\documentclass[useAMS,usenatbib]{mn2e}
\usepackage{epsfig}
\def\reff@jnl#1{{\rm#1\/}}

\def\aj{\reff@jnl{AJ}}                  
\def\araa{\reff@jnl{ARA\&A}}            
\def\apj{\reff@jnl{ApJ}}                
\def\apjl{\reff@jnl{ApJ}}               
\def\apjs{\reff@jnl{ApJS}}              
\def\apss{\reff@jnl{Ap\&SS}}            
\def\aap{\reff@jnl{A\&A}}               
\def\aapr{\reff@jnl{A\&A~Rev.}}         
\def\aaps{\reff@jnl{A\&AS}}             
\def\baas{\reff@jnl{BAAS}}              
\def\jrasc{\reff@jnl{JRASC}}            
\def\memras{\reff@jnl{MmRAS}}           
\def\mnras{\reff@jnl{MNRAS}}            
\def\physrep{\reff@jnl{Phys.Rep.}}
\def\pra{\reff@jnl{Phys.Rev.A}}         
\def\prb{\reff@jnl{Phys.Rev.B}}         
\def\prc{\reff@jnl{Phys.Rev.C}}         
\def\prd{\reff@jnl{Phys.Rev.D}}         
\def\prl{\reff@jnl{Phys.Rev.Lett}}      
\def\pasp{\reff@jnl{PASP}}              
\def\pasj{\reff@jnl{PASJ}}              
\def\skytel{\reff@jnl{S\&T}}            
\def\solphys{\reff@jnl{Solar~Phys.}}    
\def\sovast{\reff@jnl{Soviet~Ast.}}     
\def\ssr{\reff@jnl{Space~Sci.Rev.}}     
\def\nat{\reff@jnl{Nature}}             

\newcommand{\hmpc}{\ensuremath{h^{-1}\mathrm{Mpc}}}

\newcommand{\ds}{\ensuremath{\Delta\Sigma}}

\newcommand{\fclust}{\ensuremath{f_\mathrm{clust}}}
\newcommand{\fbcg}{\ensuremath{f_\mathrm{BCG}}}

\title[Halo masses of AGN]
{Halo masses for optically-selected and for radio-loud AGN from clustering 
and galaxy-galaxy lensing}
\author[Mandelbaum et al.]
{
Rachel Mandelbaum$^{1}$\thanks{E-mail: rmandelb@ias.edu, Hubble Fellow},
Cheng Li$^{2,3}$\thanks{E-mail: leech@mpa-garching.mpg.de},
Guinevere Kauffmann$^{2}$\thanks{E-Mail: gamk@mpa-garching.mpg.de},
Simon D.~M. White$^{2}$\\
$^{1}$Institute for Advanced Study, 
     Einstein Drive, Princeton NJ 08540, USA\\
$^{2}$Max-Planck-Institute for Astrophysics,
     Karl-Schwarzschild-Str. 1, D-85741 Garching, Germany \\
$^{3}$MPA/SHAO Joint Center for Astrophysical Cosmology 
     at Shanghai Astronomical Observatory, 
     Nandan Road 80, Shanghai 200030, China
}

\begin{document}

\date{Accepted ........ Received ........; in original form ........}

\pagerange{\pageref{firstpage}--\pageref{lastpage}} \pubyear{2008}

\maketitle

\label{firstpage}

\begin{abstract}
We compute two-point correlation functions and measure the shear
signal due to galaxy-galaxy lensing for 80~000 optically identified
and 5~700 radio-loud AGN from Data Release 4 (DR4) of the Sloan
Digital Sky Survey. Halo occupation models are used to estimate halo
masses and satellite fractions for these two types of AGN.  The large
sample size allows us to separate AGN according to the stellar mass of
their host galaxies. We study how the halo masses of optical and radio
AGN differ from those of the parent population at fixed $M_\ast$. Halo
masses deduced from clustering and from lensing agree satisfactorily.
Radio AGN are found in more massive halos than optical AGN: in our
samples their mean halo masses are $1.6\times 10^{13}$ and
$8\times10^{11} h^{-1} M_{\odot}$, respectively.  Optical AGN follow
the same relation between stellar mass and halo mass as galaxies
selected without regard to nuclear properties, but radio-loud AGN
deviate significantly from this relation. The dark matter halos of
radio-loud AGN are about twice as massive as those of control galaxies
of the same stellar mass. This boost is independent of radio
luminosity, and persists even when our analysis is restricted to field
galaxies. The large-scale gaseous environment of the galaxy clearly
plays a crucial role in producing observable radio emission.  The dark
matter halo masses that we derive for the AGN in our two samples are
in good agreement with recent models in which feedback from radio AGN
becomes dominant in halos where gas cools quasi-statically.
\end{abstract}

\begin{keywords}
galaxies: active -- galaxies: haloes -- galaxies: formation --
gravitational lensing -- dark matter --large-scale structure of Universe
\end{keywords}

\section{Introduction}\label{S:introduction}

It is now  widely  accepted that galaxies   form  by the  cooling  and
condensation of baryons within    a merging hierarchy of   dark matter
halos  \citep{White-Rees-78}.    Processes other   than   cooling also
influence   the  relationship  between    galaxies and   their  halos.
``Feedback,'' both from supernova explosions and from energy liberated
during the  accretion of material  onto  a central  supermassive black
hole, is currently under  considerable scrutiny and debate, because it
is believed to play  a very important  role in regulating the fraction
of available baryons that end up in galaxies.

Over  the past  few  years, considerable effort   has been  devoted to
obtaining quantitative    constraints  on   the  relationship  between
galaxies and  their dark matter  halos  using a variety  of  different
methods.  Models that  describe  the evolved,  non-linear  dark matter
distribution in terms of its halo building blocks
\citep[so-called halo models, e.g.][]{Peacock-Smith-00, Seljak-00, 
Berlind-Weinberg-02, Cooray-Sheth-02,  Yang-03} or  direct      N-body
simulations              \citep[e.g.][]{Kauffmann-Nusser-Steinmetz-97,
Jing-Mo-Boerner-98, Kauffmann-99, Benson-00a, Yang-03}, can be used in
conjunction with the  measured  clustering  amplitude of galaxies   to
constrain the relationship between  the galaxies and their host halos.
These  constraints should be regarded as  indirect, in part because of
the need for a cosmology-dependent conversion from galaxy bias to halo
mass.

Weak lensing around  galaxies (or galaxy-galaxy lensing, hereafter g-g
lensing)  provides  a {\it direct}   probe of   the dark  matter  that
surrounds galaxies   \citep[for a review, see][]{2001PhR...340..291B}.
Gravitational   lensing  induces  tangential  shear   distortions   of
background galaxies  around   foreground  galaxies,   allowing  direct
measurement of the galaxy-mass  correlation function  around galaxies.
The individual distortions  are small (of order  0.1 per cent), but by
averaging  over all foreground  galaxies within a  given subsample, we
obtain high signal  to noise in   the shear as  a  function of angular
separation from the galaxy.  If we  know the lens redshifts, the shear
signal can be  related to the projected  mass density as a function of
proper distance from the galaxy.  Thus we can  observe the averaged DM
distribution around any given galaxy sample.

These   techniques  have  been applied   to   study how galaxies  with
different properties   such   as luminosity,   stellar  mass,  colour,
spectral  type and morphology populate dark  matter halos of different
masses
\citep{2005ApJ...635...73H,2006MNRAS.371L..60H,2006MNRAS.368..715M}. The
derived relations serve as  important constraints on models of  galaxy
formation,  but they do not  directly constrain the physical processes
that are responsible for creating the relations in the first place.

For example, \citet{Yang-05} use the occupation  statistics of a group
catalogue to  show that the mean luminosity  of  halo central galaxies
scales with halo mass as $L_c \propto  M^{2/3}$ for halos less massive
than   $10^{13} h^{-1}M_{\odot}$ ,  and  flattens to  a much shallower
relation ($L_c  \propto M^{1/4}$) for  more massive halos. It has been
proposed that  this characteristic   scale   reflects the imprint   of
feedback from radio-loud AGN, which heat the gas  in massive halos and
prevent it from cooling and condensing onto the central galaxy. In the
models of \citet{Croton-06} and  \citet{Bower-06}, this ``radio mode''
feedback operates in halos with masses greater than $\sim 3
\times 10^{11} h^{-1}M_{\odot}$, where cooling times are  long compared to
the free-fall time and gas  cools quasi-hydrostatically.  On the other
hand,   the  work  of    \citet{Springel-DiMatteo-Hernquist-05}    and
\citet{Hopkins-05a, Hopkins-05b} has focused  on the role of optically
luminous AGN in expelling gas from galaxies and regulating the rate at
which they are  able to form stars.  In these
models, the major growth phases  of black holes and the triggering  of
optically luminous AGN occur when two galaxies that contain sufficient
cold gas merge  with each other. The triggering  thus does not  depend
{\it directly} on the mass of surrounding dark matter halo.

In  order  to constrain  the   importance  of  these processes, it  is
important  to understand  how the AGN  themselves are   related to the
surrounding dark  matter  distribution.   The  large-scale  clustering
amplitude of luminous quasars has been  accurately measured using tens
of thousands of such objects drawn from  the 2dF and Sloan Digital Sky
Survey.  \citet{Croom-05}  use a sample based on the 2dF to conclude
that quasars  from $0.5<z<2.5$ inhabit 
dark matter halos with a characteristic mass of $\sim 3 \times 10^{12}
h^{-1}M_{\odot}$, and that this mass does not depend on redshift or on
quasar luminosity.  \cite{2007ApJ...658...85M} use a
photometrically-identified quasar sample from SDSS over a similar redshift range
to estimate a typical mass of $\sim 5\times 10^{12}h^{-1}M_{\odot}$,
without a robust detection of luminosity-dependent bias at fixed
redshift (for which there is only a marginal detection in the 2dF
sample, \citealt{2006MNRAS.371.1824P}).  Several studies
\citep{2006AJ....131....1H,2007ApJ...658...99M,2008ApJ...678..635M}
have also 
probed the very small-scale clustering of quasars, using pairs of
binary quasars, which are a probe of how the local environment affects
quasar activity, though no clear consensus arises from these studies.
The \citet{Croom-05} results about halo masses were recently extended
to redshifts  $z>3$ by \citet{Shen-07} and $z<0.6$ by
\citet{2008arXiv0802.2105P}.   
The lack of evolution of the halo
masses from the $\sim 10^{12}h^{-1}M_{\odot}$ scale is all the more
remarkable considering that the nonlinear mass evolves by well over a
factor of ten over the full redshift range probed by these studies.

Clustering measurements of radio-loud  AGN have been considerably less
accurate because many fewer  redshifts have been available. Studies of
the  angular clustering  of  radio sources  drawn  from the  wide-area
surveys such as the NRAO VLA  Sky Survey (NVSS) or the Faint Images of
the  Radio   Sky  at  Twenty-centimetres  (FIRST)   survey  show  that
radio-loud AGN are considerably  more strongly clustered than quasars.
The  estimated  halo masses  are  typically  around $10^{13}  -10^{14}
M_{\odot}$ \citep{Overzier-03}.  \citet{Magliocchetti-04} computed the
redshift-space correlation function for  820 nearby radio sources with
redshifts  from the  2dF, and  derived  a characteristic  halo mass  of
$10^{13.4}  M_{\odot}$.  No  difference  was found  in the  clustering
properties of AGN with different radio luminosities.

This paper focuses on a sample of 80~000 optically identified AGN and
5~700 radio-loud AGN drawn from the Data Release 4 (DR4) of the Sloan
Digital Sky Survey.  We compute two-point correlation functions and
measure the shear signal due to g-g lensing for these two samples.
The large sample size allows us to split the AGN into different bins
in stellar mass and study how the dark matter halo masses of optical
and radio AGN differ at a fixed value of $M_*$.  Nearby radio-loud AGN
have been shown to have significantly higher stellar masses than
optically-identified AGN at the same redshift \citep{Best-05a}, so it
is important to understand whether the derived halo masses simply
track this difference or whether the dark matter halo affects either
the ability of an accreting black hole to produce a jet or the
detectability of the jet at radio wavelengths.

We also create control samples  of non-AGN that are matched in stellar
mass,  redshift and  morphology, and  we use  these control  samples to
investigate  whether the halo  masses of  active galaxies  differ from
those of their counterparts chosen irrespective of their
level of nuclear activity.  Finally, the fact
that the clustering  and weak lensing analyses are  carried out on the
same  set of galaxies  allows us  to evaluate  the consistency  of our
constraints on dark matter halo mass obtained using the two methods.

We begin by outlining  the theory  behind  the lensing and  clustering
measurements in section~\ref{S:theory}.    We then  describe  the data
used  for   the     analysis,   and   the    analysis  procedure,   in
section~\ref{S:data}.  The results for optical  and radio-loud AGN are
presented in section~\ref{S:results}, including lensing and clustering
separately, followed  by a joint  analysis.  We then summarize the key
results and discuss   their implications  in  sections~\ref{S:summary}
and~\ref{S:implications}, respectively.

\section{Theory}\label{S:theory}

\subsection{Galaxy-galaxy lensing}\label{S:gglensing}

Galaxy-galaxy  weak  lensing  provides  a  simple  way  to  probe  the
connection  between galaxies  and matter  via  their cross-correlation
function
\begin{equation}
\xi_{gm}(\vec{r}) = \langle \delta_g (\vec{x})
\delta_{m}(\vec{x}+\vec{r})\rangle 
\end{equation}
where $\delta_g$  and $\delta_{m}$  are overdensities of  galaxies and
matter, respectively, and in practice the mean is taken over some
survey volume (in theory it is the average over a whole distribution,
but we can only estimate its value using the fixed volumes that are available
in reality).  We will interchangably express correlation functions as
functions of vectors ($\vec{r}$) and scalars ($r$) because of the
assumption of statistical isotropy.

This cross-correlation  can be related  to the  
projected surface density
\begin{equation}\label{E:sigmar}
\Sigma(R) = \overline{\rho} \int \left[1+\xi_{gm}\left(\sqrt{R^2 + \chi^2}\right)\right] d\chi
\end{equation}
(for $r^2=R^2+\chi^2$), where we ignore the radial window, which is
much broader than the typical extent of the lens.  This surface
density  is  then related  to the  observable
quantity for lensing,
\begin{equation}\label{E:ds}
\ds(R) = \gamma_t(R) \Sigma_c= \overline{\Sigma}(<R) - \Sigma(R), 
\end{equation}
where the second relation is  true only for a matter distribution that
is axisymmetric along the line of sight.  This observable quantity can
be  expressed  as the  product  of  two  factors, a  tangential  shear
$\gamma_t$ and a geometric factor, the critical surface density 
\begin{equation}\label{E:sigmacrit}
\Sigma_c = \frac{c^2}{4\pi G} \frac{D_S}{D_L D_{LS}(1+z_L)^2}
\end{equation}
where $D_L$ and $D_S$ are angular diameter distances to the lens and
source, $D_{LS}$ is the angular diameter distance between the lens
and source, and the factor of $(1+z_L)^{-2}$ arises due to our use of
comoving coordinates.  For a given lens redshift,
$\Sigma_c^{-1}$ rises from zero at $z_s = z_L$ to an asymptotic value
at $z_s \gg z_L$; that asymptotic value is an increasing function of
lens redshift.  

In practice, we  measure the g-g weak lensing  signal around a stacked
sample of  lenses to obtain the average $\Delta\Sigma(R)$ for  the whole
sample.  This stacked lensing signal  can be split into two terms that
dominate on  different scales.  The  1-halo or Poisson term,  which is
determined  by the  dark matter  halo in  which the  galaxy lives,
dominates on  scales typically below $\sim  1h^{-1}$Mpc. The halo-halo
term, which is determined by correlations between the galaxy and other
dark matter halos, dominates on larger scales.  The 1-halo term can be
further  split into  two contributions.   For central  galaxies, which
reside at the peak density of a dark matter halo that is not contained
within  another halo  (a  host  halo), the  Poisson term is  simply
determined by  the matter  density of that  host halo,  $\rho(r)$.  For
satellite galaxies, which reside in  dark matter subhalos, there is a
contribution from the density of the dark matter subhalo, but there is
also  a term  on  hundreds  of  kiloparsec scales  due to  the
cross-correlation between the galaxy  position and the {\em host} dark
matter halo. Consequently, the lensing signal on $<\sim 0.3
h^{-1}$Mpc\ 
scales  tells  us about  the  dark matter  halo  in  which the  galaxy
resides; the  signal from $\sim  0.3$ -- $1$\hmpc\  reveals the
local environment of the galaxy; and the signal on larger scales
indicates the large-scale correlations of the galaxy sample. 

We  interpret the  lensing  signal statistically  using  a halo  model,
which  allows us to
determine both  the typical halo mass $M_{cent}$  for central galaxies
in our  galaxy sample, and  also the satellite fraction  $\alpha$ (the
fraction of the sample located in subhalos within some more massive
host dark matter halo).  In
this simple formulation of the  halo model, we assume that all central
galaxies in  our sample have a  single halo mass $M_{cent}$, and that all
satellite galaxies are distributed in halos with $M>3M_{cent}$ with
the number in a halo of given mass above this threshold $\propto M$.
Tests  of this halo 
model  formulation using  the lensing  signal from  N-body simulations
\citep{2005MNRAS.362.1451M}  clearly  indicate  that the  best-fitting
$M_{cent}$ and  $\alpha$ recover the  true values to within  $\sim 10$
per cent,  provided that  the distribution of  central halo  masses is
relatively  narrow (FWHM  typically a  factor of  6 or  less).   For a
broader  distribution   of  central  halo   masses,  the  best-fitting
$M_{cent}$ lies between  the median and the mean  of the distribution,
but may  differ from either one  by as much  as a factor of  two.  For
more   details  of   this  halo   model  and   its   assumptions,  see
\cite{2005MNRAS.362.1451M}.   In this  work,  we use  this halo  model
without  applying any  correction  for the  (unknown) scatter  between
galaxy stellar mass and dark matter halo mass, but we will discuss the
extent to  which our assumptions about  the width of  the central halo
mass distribution are likely to be correct.

\subsection{Galaxy clustering}\label{S:gclustering}

The clustering of galaxies is usually quantified using  the two-point
correlation function \citep[2PCF, e.g.][]{Peebles-80}, defined by
\begin{equation}
d P_{12} = \bar{n}^2[1+\xi(\vec{r})]dV_1dV_2. 
\end{equation}
Here $\bar{n}$ is the mean number  density of galaxies, and $dV_1$ and
$dV_2$  are the volumes of two  infinitesimally small spheres centered
at      $\vec{x}_1$     and     $\vec{x}_2$    with     distance    of
$\vec{r}=\vec{x}_2-\vec{x}_1$.  By definition, $d P_{12}$ is the joint
probability that a galaxy lies in each of the spheres, and so the 2PCF
$\xi(r)$ represents  the  excess probability  of  finding two galaxies
separated by  a distance $\vec{r}$, compared  with the result obtained
for a uniform random  distribution.  If $\xi(r)>0$, then  galaxies are
said  to  be  clustered.  In   galaxy  redshift surveys,  the 2PCF  is
measured   in  redshift space and    usually expressed as functions of
separations perpendicular ($r_p$) and parallel ($\pi$)  to the line of
sight.  In many cases,  the  projected two-point correlation function,
$w_p(r_p)$, is  the more useful  quantity, because  it does not suffer
from redshift-space   distortions,   and is  thus   directly  related to  the
real-space correlation function.   The 2PCF is  also simple to compute
and can be  easily compared with  the predictions of  theoretical
models.

The amplitude of the correlation function on scales  larger than a few
Mpc provides a direct  measure of the  mass  of the dark  matter halos
that host the galaxies through the halo mass - bias relation. As shown
in \citet{Li-08a, Li-08b}, the  amplitude of the  correlation function
on scales $\la$  100 kpc  can serve  as a probe  of physical processes
such as mergers and interactions. On intermediate scales, the shape of
the correlation function is sensitive  to how galaxies are distributed
{\em within} their dark matter halos.

The clustering signal must also be interpreted using some form of halo
model.  Following the approach adopted in our previous work, we
interpret clustering results using the models of \citet{Li-06a} and
\citet{Wang-06} which are based on direct N-body simulations.  We have
constructed a set of 100 mock galaxy catalogues from the Millennium
Simulation \citep{Springel-05} with exactly the same observational
selection effects as the SDSS DR4.  The Millennium Simulation uses
$10^{10}$ particles to follow the dark matter distribution in a cubic
region 500$h^{-1}$ Mpc on a side.  The cosmological parameters assumed
are $\Omega_m=0.25$, $\Omega_\Lambda=0.75$, $\sigma_8=0.9$ and
$h=0.73$.  We adopted the positions and velocities of the galaxies
given in the catalogue of \citet{Croton-06}, who implemented a
semi-analytic model in order to track the formation and evolution of
galaxies in the simulation.  Physical properties of the galaxies, such
as stellar masses and AGN status, are not taken from the semi-analytic
model, however, but instead are assigned to each model galaxy using
parametrized functions.  The main such function relates the stellar
mass of the galaxy to the mass of the halo at the epoch when the
galaxy was last the central dominant object in its own halo, including
scatter in that relation.  Tests
have shown that this procedure allows us to match accurately both the
stellar mass function of SDSS galaxies and the shape and amplitude of
their two-point correlations as a function of stellar mass
\citep{Li-06a, Wang-06}.

In \citet[][hereafter L06]{Li-06c}, we adapted this halo model to
interpret the clustering of optically-identified AGN.  In that paper,
we computed the correlation functions of AGN and of control samples of
{\em inactive} galaxies that had the same redshift and stellar mass
distribution as the active galaxies.  We found that on scales between
100 kpc and 1 Mpc, AGN are clustered more weakly than the inactive
sample.  We then introduced a simple model in which the probability of
a galaxy of given stellar mass to be an AGN is enhanced if it is the
central galaxy of its own halo, and showed that this model could
provide a good fit to the data. In the best-fitting model, 84 per cent
of all optical AGN are located at the centres of their own dark matter
halos (i.e., $f_{cen}=0.84$), whereas this is true for only 73 per
cent of inactive galaxies.  

We emphasize that while in principle, it
would be easy to assume that $\alpha$ derived from the lensing
analysis is simply $1-f_{cen}$ from the clustering analysis, the
relationship is not as straightforward as this.  The clustering
analysis starts from the assumption that the AGN and control samples
derive from a parent population which matches the statistical
properties of galaxies as a function of stellar mass (i.e., the
stellar mass function and the clustering as a function of mass).  Any
halo model parameters such as $f_{cen}$ are then introduced as a way
of matching the observed AGN clustering by modifying the probability
that each galaxy in the parent population is an AGN.  In
contrast, the lensing analysis does not assume that the AGN and
control samples stem from identical parent populations.  In future
analyses with larger datasets, we will jointly model clustering and
lensing with the same assumptions at the outset; here, we simply use
pre-existing analysis pipelines that output quantities that should be
reasonably (but not exactly) comparable.  Even in the absence of a
completely unified approach to modeling, there are many valuable
conclusions that can be drawn from the lensing and
clustering signals (e.g., if both the clustering and lensing signals for a
particular sample are comparable to the signals for the controls, or
if they are both quite different from the signals for the controls).

\section{Data and signal measures}\label{S:data}

\subsection{Overview of SDSS}
The data used here are obtained from the SDSS
\citep{2000AJ....120.1579Y}, an ongoing survey to image roughly $\pi$
steradians of the sky, and follow up approximately one million of the
detected objects spectroscopically \citep{2001AJ....122.2267E,
2002AJ....123.2945R,2002AJ....124.1810S}.  The imaging is carried out
by drift-scanning the sky in photometric conditions
\citep{2001AJ....122.2129H, 2004AN....325..583I}, in five bands
($ugriz$) \citep{1996AJ....111.1748F, 2002AJ....123.2121S} using a
specially-designed wide-field camera
\citep{1998AJ....116.3040G}. These imaging data are used to create the
source catalogue that we use in this paper. In addition, objects are
targeted for spectroscopy using these data \citep{2003AJ....125.2276B}
and are observed with a 640-fiber spectrograph on the same telescope
\citep{2006AJ....131.2332G}.  All of these data are processed by
completely automated pipelines that detect and measure photometric
properties of objects, and astrometrically calibrate the data
\citep{2001ASPC..238..269L,
2003AJ....125.1559P,2006AN....327..821T}. The SDSS has had seven major
data releases \citep{2002AJ....123..485S, 2003AJ....126.2081A,
2004AJ....128..502A, 2005AJ....129.1755A, 2004AJ....128.2577F,
2006ApJS..162...38A, 2007ApJS..172..634A,2008ApJS..175..297A}. In this
paper we use data from the fourth of these
releases\citep[DR4;][]{2006ApJS..162...38A}.

\subsection{The AGN and control samples}

\subsubsection{Optically-identified AGN}\label{SS:optagn}

The sample of optically-identified AGN is the same as that analyzed in
L06 and \citet{Li-08b}, in which the clustering of AGN on a variety of
different scales was studied.

The base sample is composed of $\sim 4\times 10^5$ objects for which
data are publicly available through DR4 and which have been
spectroscopically confirmed as galaxies with $r$-band magnitudes in
the range $14.5<r<17.6$, redshifts in the range $0.01<z<0.3$, and
absolute magnitudes in the range $-23 < M_{0.1_{r}}< -17$.  Here $r$
is the $r$-band Petrosian apparent magnitude corrected for foreground
extinction, and $M_{^{0.1}r}$ is the $r$-band absolute magnitude
corrected to its $z=0.1$ value using the $k$-correct code of
\cite{Blanton-03b} and the luminosity evolution model of
\cite{Blanton-03c}.  A sample of $\sim 80,000$ AGN are selected from
the subset of these galaxies with $S/N>3$ in the four emission lines
[O {\sc iii}]$\lambda$5007, H$\beta$, [N {\sc ii}]$\lambda$6583 and
H$\alpha$, following the criteria proposed by \citet{Kauffmann-03}.
In the following analysis, we occasionally divide our sample into
``weak'' and ``powerful'' AGN using the quantity $L$[OIII]$/M_{bh}$,
where $L$[OIII] is the extinction-corrected [OIII] line luminosity of
the AGN and M$_{bh}$ is the black hole mass estimated from the
velocity dispersion of the galaxy using the relation given in
\citet{Tremaine-02}.  As discussed in \citet {Heckman-04}, this
quantity can be viewed as a measure of the accretion rate onto the
black hole relative to the Eddington rate.

We have also constructed two sets of 20 different control samples from
the full parent  sample of galaxies by matching  a number  of physical
parameters  regardless of nuclear activity.  For   the first set, four
physical parameters   are  matched:  redshift   ($z$), stellar    mass
($M_\ast$),   concentration  ($R_{90}/R_{50}$),  and  stellar velocity
dispersion ($\sigma_\ast$).  For  the second set,  the 4000 \AA\ break
strength  ($D_{4000}$) is also  matched.   The matching tolerances are
$\Delta     cz<500$     km    s$^{-1}$,      $\Delta\log  M_\ast<0.1$,
$\Delta\sigma_\ast<20$ km   s$^{-1}$,  $\Delta R_{90}/R_{50}<0.1$  and
$\Delta D_{4000}<0.05$.\footnote{This  procedure  is identical to that
in  L06, except that  the control galaxies are  selected from the full
parent sample, rather than from the subset of inactive galaxies.}  In
the first case, 28 per cent of the control galaxies are also included
in the AGN sample; in the second case, the overlap fraction is higher,
37 per cent.  In both cases, the match fraction is a slightly
increasing function of stellar mass.

\subsubsection{Radio-loud AGN}

The  National  Radio
Astronomy  Observatory (NRAO)  Very  Large  Array   (VLA) Sky   Survey
\citep[NVSS;][]{Condon-98} and  the Faint Images of   the Radio Sky at
Twenty centimeters (FIRST)  survey \citep{Becker-95} are two radio
surveys that have been carried out in recent years using the 
VLA radio synthesis telescope at a frequency of 1.4 GHz.
\citet{Best-05a} identified radio-emitting galaxies within the
main  spectroscopic  sample  of  the SDSS  data   release  2  (DR2) by
comparing these galaxies  with  a combination  of these two  surveys.            
The use of two radio  surveys allowed a radio  sample to be  constructed
that was  both  reasonably  complete  ($\sim$95 per cent)   and highly
reliable  (it is estimated that  $\sim$99 per cent of the sources
in the  catalogue are    genuine  radio galaxies  rather  than   false
matches). In this paper, we use an  updated catalogue of 5~712
radio galaxies based on the SDSS data release 4 (DR4).
The sample spans the same redshift range as the
 sample of optically-selected AGN, and the radio luminosities
of the AGN range from $10^{23}$ to $10^{26}$ W Hz$^{-1}$ (i.e.,
they are mainly FRI type systems).

We have used  the parent galaxy catalogue to  construct a  set of five
control samples that are closely matched in redshift, stellar mass and
stellar velocity dispersion. The matching  tolerances are the same  as
used  to construct   the optical AGN   control   samples.  We  do  not
additionally match  in D$_{4000}$, because  almost all galaxies at the
relevant stellar masses are red, with a strong 4000 \AA\ break.  Of
the control sample, 10.3 per cent of the galaxies are radio-loud AGN;
for our higher stellar mass subsample ($\log{(M_*/M_{\odot})}\ge 11.44$),
the radio-loud AGN fraction is 16 per cent, whereas for the lower
$M_*$ subsample, it is 7 per cent.

\begin{figure*}
\includegraphics[width=6in,angle=0]{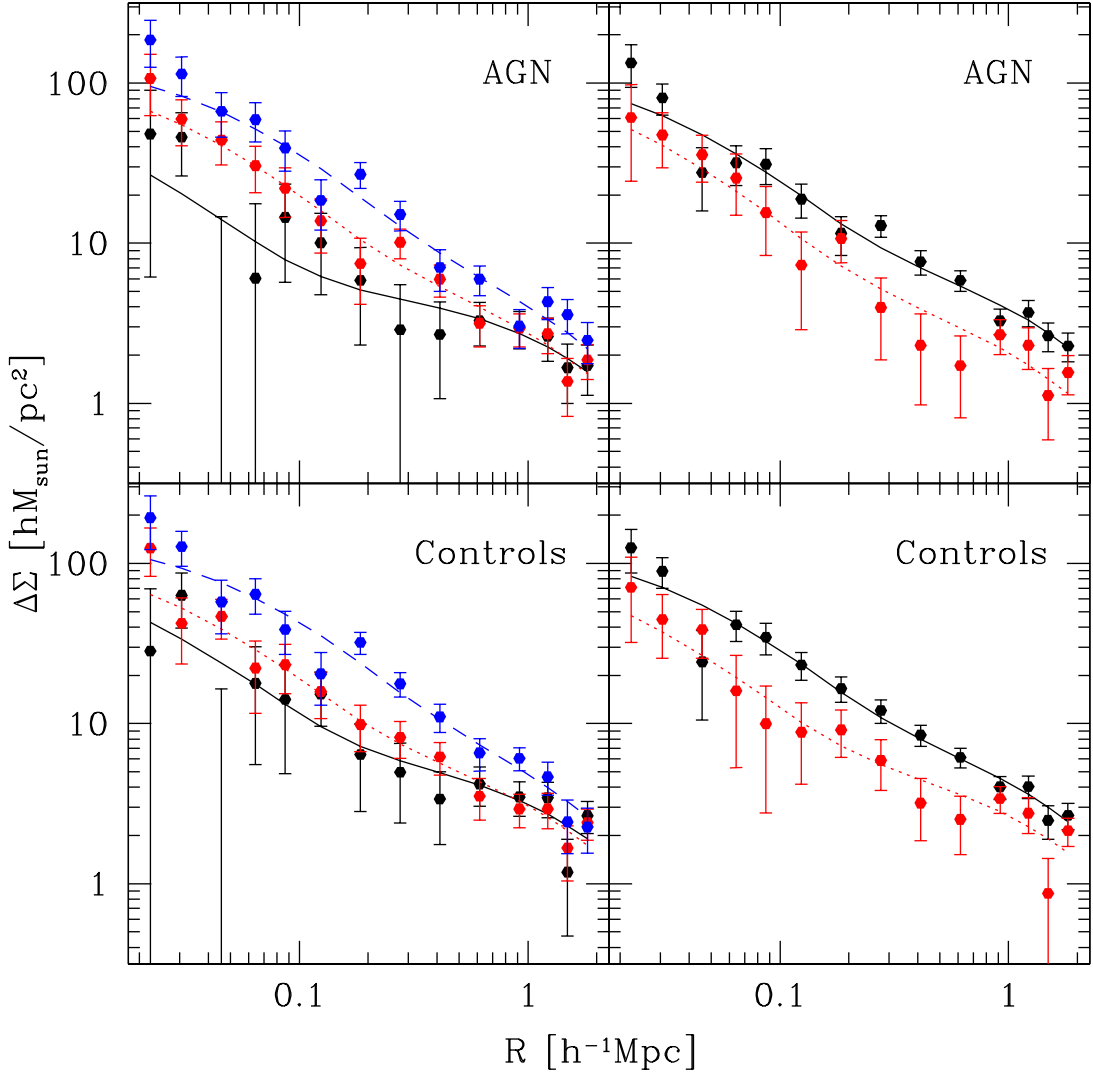}
\caption{\label{F:gg-optagn}The   galaxy-galaxy  lensing   signal  for
optical AGN (top) and control galaxies (bottom) as a function of
transverse  separation.   Points  are  the  data, and  lines  are  the
best-fitting halo  models.  The left  panels show the sample  split by
stellar   mass    as   follows:   $\log{M_*}<10.6$    (black   solid),
$10.6\le\log{M_*}<11$ (red  dotted), $\log{M_*}\ge 11$  (blue dashed).
The right panels show the  sample split by $L$[OIII]$/M_{bh}$ into the
lower half of the sample (black solid) and upper half (red
dotted).}
\end{figure*}

\subsection{Lensing analysis}

\subsubsection{Lensing source catalogue}

The source sample  used for the lensing  analysis is the same as  that
originally   described in  \cite{2005MNRAS.361.1287M}.    This  source
sample  includes over 30 million galaxies  from  the SDSS imaging data
with $r$-band    model   magnitude brighter than 21.8,      with shape
measurements  obtained   using  the REGLENS   pipeline,  including PSF
correction done via re-Gaussianization \citep{2003MNRAS.343..459H} and
with  cuts designed  to avoid  various shear   calibration biases.  In
addition  to  these, there are  also  uncertainties due to photometric
redshifts and/or  redshift distributions   of background galaxies,  as
well as due to  other issues affecting the  calibration of the lensing
signal,  such   as   the sky   subtraction  uncertainties,   intrinsic
alignments,   magnification    bias,    star-galaxy   separation,  and
seeing-dependent systematics.  The overall calibration uncertainty was
estimated to be eight per cent \citep{2005MNRAS.361.1287M}, though the
redshift calibration component   of this systematic  error budget  has
recently been decreased due to the  availability of more spectroscopic
data  \citep{2008MNRAS.386..781M}.    With a total  estimated  lensing
calibration  uncertainty of  $\sim  5$  per cent,  this  systematic is
subdominant  compared to the  statistical error and to the uncertainty
derived from the model used to interpret the lensing signal.

\subsubsection{Lensing signal computation}

Here we  briefly describe the  computation of the  lensing signal; for
more  detail,   see~\cite{2005MNRAS.361.1287M}.   For   each  lens, we
identify  sources  within 46  logarithmically-spaced annuli around the
lens (in comoving  transverse   separation) from 20 $h^{-1}$kpc   to 2
$h^{-1}$Mpc.  The tangential ellipticity of the source relative to the
lens is measured, in order to estimate the tangential shear. 
Lens-source pairs are assigned  weights according to the 
error on the shape measurement via
\begin{equation}
w_{ls} = \frac{\Sigma_c^{-2}}{\sigma_s^2 + \sigma_{SN}^2}
\end{equation}
where $\sigma_{SN}^2$ is the intrinsic shape noise and $\sigma_s$ is
the measurement error on the source galaxy ellipticity.  The factor of
$\Sigma_c^{-2}$ optimally weights the signal by the noise in
$\Delta\Sigma$ rather than in the shear.

Once we have computed these weights, we compute the lensing signal in
each radial bin  as
a summation over lens-source pairs via:
\begin{equation}
\ds(R) = \frac{\sum_{ls} w_{ls} e_t^{(ls)} \Sigma_c}{2 {\cal
    R}\sum_{ls} w_{ls}} 
\end{equation}
where the factor of 2 and the shear responsivity ${\cal R}$ relate our
definition of ellipticity to the
shear, using the formalism in \cite{2002AJ....123..583B}.  In practice, ${\cal
  R}\approx 1-e_{rms}^2 \approx 0.86$.

There  are several  additional  procedures  that   must be  done  when
computing     the       signal     (for   more          detail,    see
~\citealt{2005MNRAS.361.1287M}).   First, the  signal computed  around
random points must be subtracted from the signal around real lenses to
eliminate contributions  from  systematic  shear.  In practice,   this
correction is negligible  for the scales used  in this work.  Second,
the    signal  must be  boosted,     i.e.     multiplied by $B(R)    =
n(R)/n_{rand}(R)$, the ratio of the number density of sources relative
to the number  around random points,  in order to account for dilution
by  sources that are physically associated  with lenses, and therefore
not lensed.

To determine errors  on the lensing signal, we  divide the survey area
into 200  bootstrap subregions, and  generate 2500 bootstrap-resampled
datasets.   These bootstrap-resampled  datasets are  also  crucial for
determining  the  statistical   significance  of  differences  between
correlated subsamples of galaxies,  because fitting the signal to the
halo model on each resampled dataset allows us to determine
how much any overlap between two galaxy samples leads  to a correlation  between the best-fitting  halo   model
parameters for the two samples. 

The lensing signal is presented in comoving coordinates, with angular
diameter distances computed assuming a flat $\Lambda$CDM universe with
$\Omega_m=0.3$ and $\Omega_{\Lambda}=0.7$.  The halo model used to
interpret the lensing signal assumes $\sigma_8=0.9$.  In the units
used, $H_0$ scales out of everything, so our results are independent
of this quantity.  The central halo mass definition for this paper is
the mass within which the spherical overdensity is $200\rho_{crit}$,
which  is roughly
35 per cent lower than the mass definition used for 
previous lensing analyses using this halo model formalism
\citep{2005MNRAS.362.1451M,2006MNRAS.368..715M}.  This change in halo
mass definition 
was made to match the mass definition for the clustering analysis.

\subsection{Clustering analysis}

\subsubsection{The reference galaxy sample}

In this paper,  the  clustering  of  AGN   (or control  galaxies)   is
quantified    by the projected   two-point  cross-correlation function
(2PCCF), $w_p(r_p)$,  which is estimated  by cross-correlating the AGN
(or  control)  samples  described above  with   a reference  sample of
galaxies.\footnote{We use   the  notation  $r_p$ for  the   transverse
separation in the  clustering analysis, and the  notation $R$  for the
same quantity in the lensing analysis.  The main reason is to maintain
notational consistency   within previous work.}  The  reference galaxies
are selected from  {\tt sample dr4} of  the New York University  Value
Added Galaxy  Catalogue (NYU-VAGC),   which  is  based on  SDSS   DR4,
publicly   available  at  http://sdss.physics.nyu.edu/vagc/,    and is
described  in detail  in  \citet{Blanton-05}.   The  reference  sample
contains 292,782 objects that are identified as galaxies from the Main
sample   and  have   $0.01\leq      z\leq 0.3$,   $14.5<r<17.6$    and
$-23<M_{^{0.1}r}<-17$. This  sample has formed  the
basis of  our recent  investigations  of the clustering  properties of
different classes of galaxies \citep[L06, ][]{Li-07,Li-08a,Li-08b}.

\subsubsection{Clustering measures}

Our  methodology for  computing  correlation functions  has also  been
described  in  detail in  our  previous  papers.   Random samples  are
constructed with the  same selection function as  the reference
sample, as described  in  detail   in  \citet{Li-06a} (but note the
slight differences mentioned here in \S\ref{SS:optagn}).   The
redshift-space 2PCCF $\xi(r_p,\pi)$  between AGN (or control galaxies)
and  the  reference sample  is  then  calculated  using the  estimator
presented  in   L06,
\begin{equation}
\xi(r_p,\pi) = \frac{N_R}{N_D} \frac{QD(r_p,\pi)}{QR(r_p,\pi)} -1,
\end{equation}
where  $r_p$ and $\pi$  are the separations perpendicular and parallel
to the line  of sight; $N_D$ and $N_R$  are the number of  galaxies in
the reference sample   and  in the  random sample, with $N_R/N_D=10$ 
throughout this paper;   $QD(r_p,\pi)$ and
$QR(r_p,\pi)$ are the cross pair   counts between AGN/control and  the
reference sample, and   between AGN/control  and  the random   sample,
respectively.  Finally, the redshift-space  projected 2PCCF $w_p(r_p)$
is  estimated by  integrating  $\xi(r_p,\pi)$ along the  line-of-sight
direction:
\begin{equation}
w_p(r_p)=\int_{-\pi_{max}}^{+\pi_{max}}\xi(r_p,\pi)d\pi=
\sum_i\xi(r_p,\pi_i)\Delta\pi_i.
\end{equation}
Here $\pi_{max}=40 h^{-1}$ Mpc, and the summation for computing 
$w_p(r_p)$ runs from $\pi_1 = -39.5 $
h$^{-1}$ Mpc to $\pi_{80} = 39.5$ h$^{-1}$ Mpc, with $\Delta\pi_i = 1$
h$^{-1}$ Mpc. We have also corrected carefully for the effect of fibre
collisions;  a description and  tests of  the  method can be found  in
L06. As will be described in more detail in
\S~\ref{S:radioAGN_gclustering}, error estimates 
come from the variance in $w_p(r_p)$ between 100 mock catalogues.

The clustering computation assumes the same flat $\Lambda$CDM universe
with $\Omega_m=0.3$,  $\Omega_{\Lambda}=0.7$ and $\sigma_8=0.9$ as for
the lensing analysis.  Our results  are presented in units of $h^{-1}$
Mpc with $h=1$.

\section{Results}\label{S:results}

\subsection{Optical AGN}\label{S:opticalAGN}

Results  and interpretation for  the galaxy  clustering signal  of the
optical AGN have  been presented in L06, and are  also briefly described in
\S\ref{S:gclustering}.   Consequently,   here  we  present   only  the
galaxy-galaxy lensing signal  and its interpretation for this  sample.
In section~\ref{S:summary}, we compare  the halo masses  and satellite
fractions estimated through lensing with the same quantities estimated
through clustering.

In Fig.~\ref{F:gg-optagn},   we show the  g-g  lensing  signal for the
optical AGN sample split by stellar mass and by the accretion rate per
unit black hole mass ($L$[OIII]$/M_{bh}$).  Results are shown for both
the AGN and the control samples; results for  the control samples with
the same  distribution of $D_{4000}$  are not shown,  because they are
statistically  consistent  with the  results for  the  control samples
where $D_{4000}$ is not matched.

There are several clear trends in this figure.  First, the g-g lensing
signal   on small scales  ($<0.3$\hmpc)  shows that  the halo mass for
central  galaxies increases  with   stellar  mass and  decreases  with
$L$[OIII]$/M_{bh}$.  This conclusion is true for  both the AGN and for
the control samples.   Second, the  g-g lensing   signal for AGN   and
controls  in a particular subsample is  quite similar; any differences
are not statistically significant.

\begin{figure}
\includegraphics[width=3.2in,angle=0]{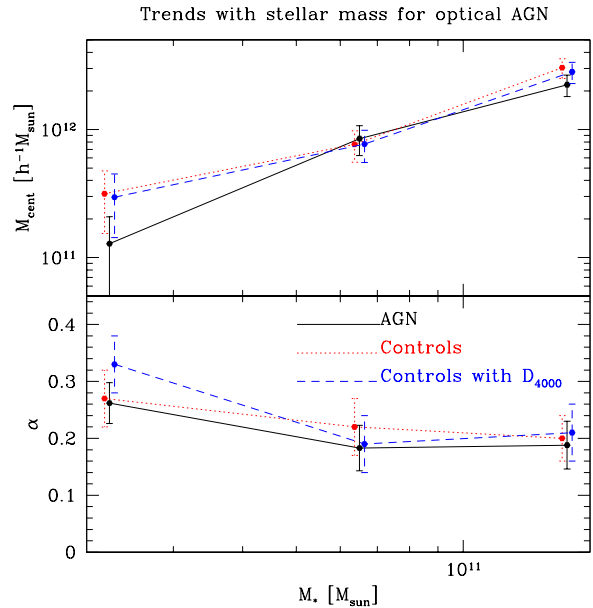}
\caption{\label{F:gg-optagn-hmfits}The best-fitting central halo
  masses $M_{cent}$ (top) and satellite fractions $\alpha$ (bottom) as
  a function of stellar mass for the optical AGN and the two control
  samples as labelled on the plot.}  
\end{figure}

We  now  consider the  halo  model  interpretation  of these  results,
represented by the   best-fitting   central halo mass   and  satellite
fraction for each    sample.  These quantities   are plotted  for  the
optical     AGN   and       the      two    control       samples   in
Fig.~\ref{F:gg-optagn-hmfits} as a   function  of stellar  mass.   For
reference,   they  are tabulated  for all    optical  AGN and  control
subsamples,   including    the  splits   by   $L$[OIII]$/M_{bh}$,   in
Table~\ref{T:gg-optagn-hmfits}.

\begin{figure}
\includegraphics[width=3.2in,angle=0]{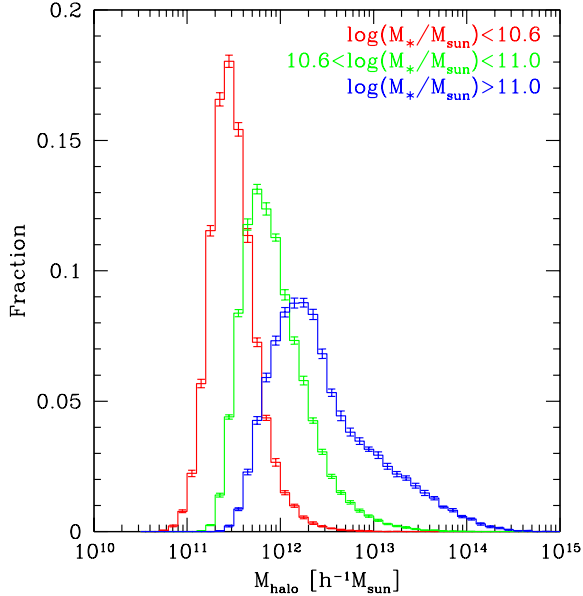}
\caption{\label{F:mcentdist-optagn}The distributions of central halo
  mass in the mock catalogues that are able to reproduce the clustering
  signal in \protect\cite{Li-06c} in our three stellar
  mass bins.}  
\end{figure}

\begin{table*}
\caption{\label{T:gg-optagn-hmfits}Best-fitting halo model parameters for
  fits to the optical AGN g-g weak lensing signal, with 68 per cent CL errors.}
\begin{tabular}{lcccccc}
 & \multicolumn{2}{c}{Optical AGN} & \multicolumn{2}{c}{Controls} &
 \multicolumn{2}{c}{Controls with $D_{4000}$} \\
Sample & $M_{cent}$ &  $\alpha$ & $M_{cent}$ &  $\alpha$ & $M_{cent}$ &  $\alpha$ \\
 & $10^{12} h^{-1}M_{\odot}$ & & $10^{12} h^{-1}M_{\odot}$ & & $10^{12} h^{-1}M_{\odot}$ & \\
\hline
\hline
$\log{M_*/M_\odot} < 10.6$, $\langle M_*/(10^{10}M_\odot)\rangle=2.3$
& $0.13\pm 0.08$ & $0.26\pm 0.04$ & $0.31\pm 0.16$ & $0.27\pm 0.05$ &
$0.29\pm 0.15$ & $0.33\pm 0.05$ \\
$10.6\le \log{M_*/M_\odot} < 11$, $\langle
M_*/(10^{10}M_\odot)\rangle=6.5\!\!\!\!\!$ & $0.86\pm 0.21$ & $0.18\pm 0.04$ &
$0.77\pm 0.20$ & $0.22\pm 0.05$ & $0.79\pm 0.21$ & $0.19\pm 0.05$ \\
$\log{M_*/M_\odot} \ge 11$, $\langle M_*/(10^{10}M_\odot)\rangle=15.4$
& $2.3\pm 0.4$ & $0.19\pm 0.04$ & $3.0\pm 0.5$ & $0.20\pm 0.04$ &
$2.8\pm 0.6$ & $0.21\pm 0.05$ \\
Lower half, $L$[OIII]$/M_{bh}$ & $1.1\pm 0.2$ & $0.25\pm 0.03$ &
$1.5\pm 0.2$ & $0.25\pm 0.03$ & $1.1\pm 0.2$ & $0.30\pm 0.03$ \\
Upper half, $L$[OIII]$/M_{bh}$ & $0.47\pm 0.13$ & $0.16\pm 0.03$ &
$0.39\pm 0.14$ & $0.22\pm 0.03$ & $0.52\pm 0.16$ & $0.17\pm 0.03$ \\
\hline
\hline
\end{tabular}
\end{table*}

A few  trends  are evident from the   plot and table.  First, for  the
samples   split   by stellar   mass,  the  differences  in  halo model
parameters for the AGN  and control samples are not  statistically
significant,  as expected  from Fig.~\ref{F:gg-optagn}.   However, the
central halo mass  shows a strong  trend with stellar mass, consistent
with the  lensing results for  the general galaxy population discussed
in \cite{2006MNRAS.368..715M}.  If  we compare against the  results in
that paper after accounting for the different halo mass definitions,
we  conclude that  for the  lower and middle  stellar mass 
bins,   the best-fitting central halo  mass  is consistent (within the
noise) with the results  for both early  and late type galaxies, which
have similar mean halo masses below  stellar masses $\sim 10^{11}M_{\odot}$.  For the
highest stellar mass  bin, our best-fitting  central halo mass is more
consistent with the results  for late-type galaxies (lower  by a factor 
of a few  than that for early-type galaxies).  This result is consistent
with the general  tendency of these  narrow-line AGN to be  associated
with galaxies with ongoing star formation.

The satellite fractions decrease   slightly from the lowest to  middle
stellar mass  bin.   Consistent  with the  results from   L06, we find
slightly  lower satellite fractions for the  optical  AGN than for the
control    samples.  Unlike for the  galaxy     clustering signal, this
difference is not statistically significant.

There  is  clearly a significant  difference  in the mean central halo
mass and satellite fraction for the  samples split at the median value
of $L$[OIII]$/M_{bh}$.  However,   this quantity is itself  correlated
with stellar mass, so some of the trend derives from that correlation.
For the lower half of the sample in  OIII luminosity, the mean stellar
mass is  $\langle M_*\rangle  =  9.1\times 10^{10}M_{\odot}$;  for the
upper half, it is  $7.3\times 10^{10}M_{\odot}$.  The results for  the
control  samples with the same  stellar mass distribution suggest that
the difference  in best-fitting central  halo masses can  be explained
solely by this difference in stellar mass distributions.

Finally,   we   discuss  the broadness  of    the  central  halo  mass
distribution.  As we have  already noted, a   broad central halo  mass
distribution would lead to the  best-fitting central halo masses being
an   overestimate of the  median mass,  and underestimate of
the mean mass.   To assess whether this may  be the  case, we use  the
halo   occupation  models     of      L06 for    optical    AGN   (see
\S\ref{S:gclustering}) which can be used to derive a central halo mass
distribution.  This plot is shown in Fig.~\ref{F:mcentdist-optagn}. As
shown, the  FWHM of the distribution  for  the two  lower stellar mass
bins is within the  factor of $\sim  6$ needed for accurate estimation
of  the   mean  central  halo mass.     However,   the distribution is
sufficiently broad for the highest  stellar mass bin that our estimate
from   the  lensing signal is likely    an underestimate of  the mean,
possibly by as much as  50 per cent.  We do  not apply a correction to
determine  the  mean  central  halo mass for   either the   AGN or the
controls   in    this bin,  since    there  is  significant systematic
uncertainty in the correction factor itself.

\subsection{Radio AGN}\label{S:radioAGN}

\subsubsection{Galaxy Clustering}\label{S:radioAGN_gclustering}
\begin{figure*}
\centerline{\psfig{figure=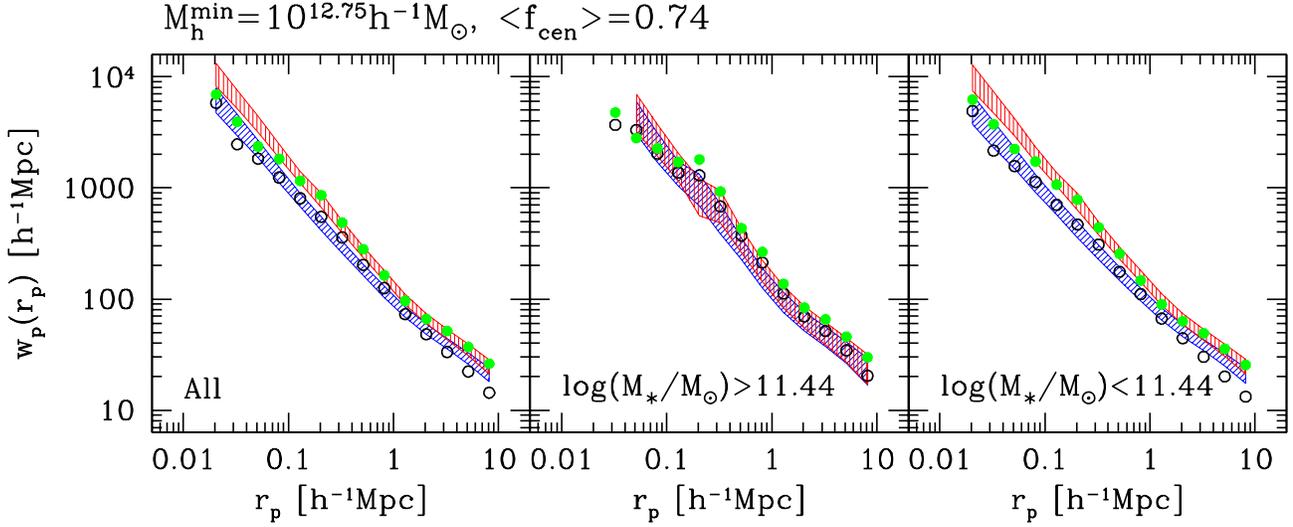,clip=true,width=\textwidth}}
\caption{Projected cross-correlation function $w_p(r_p)$ for radio AGN
(green filled circles)  in different stellar  mass ranges  compared to
results for control samples  selected without regard to AGN properties
(black open symbols).  Results for the best-fitting model 
are indicated as red (AGN) and blue (controls)  shaded regions, where  the width of the
shaded regions corresponds to  the $1-\sigma$ variance between 200 mock
catalogues.  Errorbars are significantly correlated ($>10$ per cent) between radial
bins above $\sim 1h^{-1}$Mpc.}
\label{fig:wrp_model_B}
\end{figure*}

\begin{figure*}
\centerline{\psfig{figure=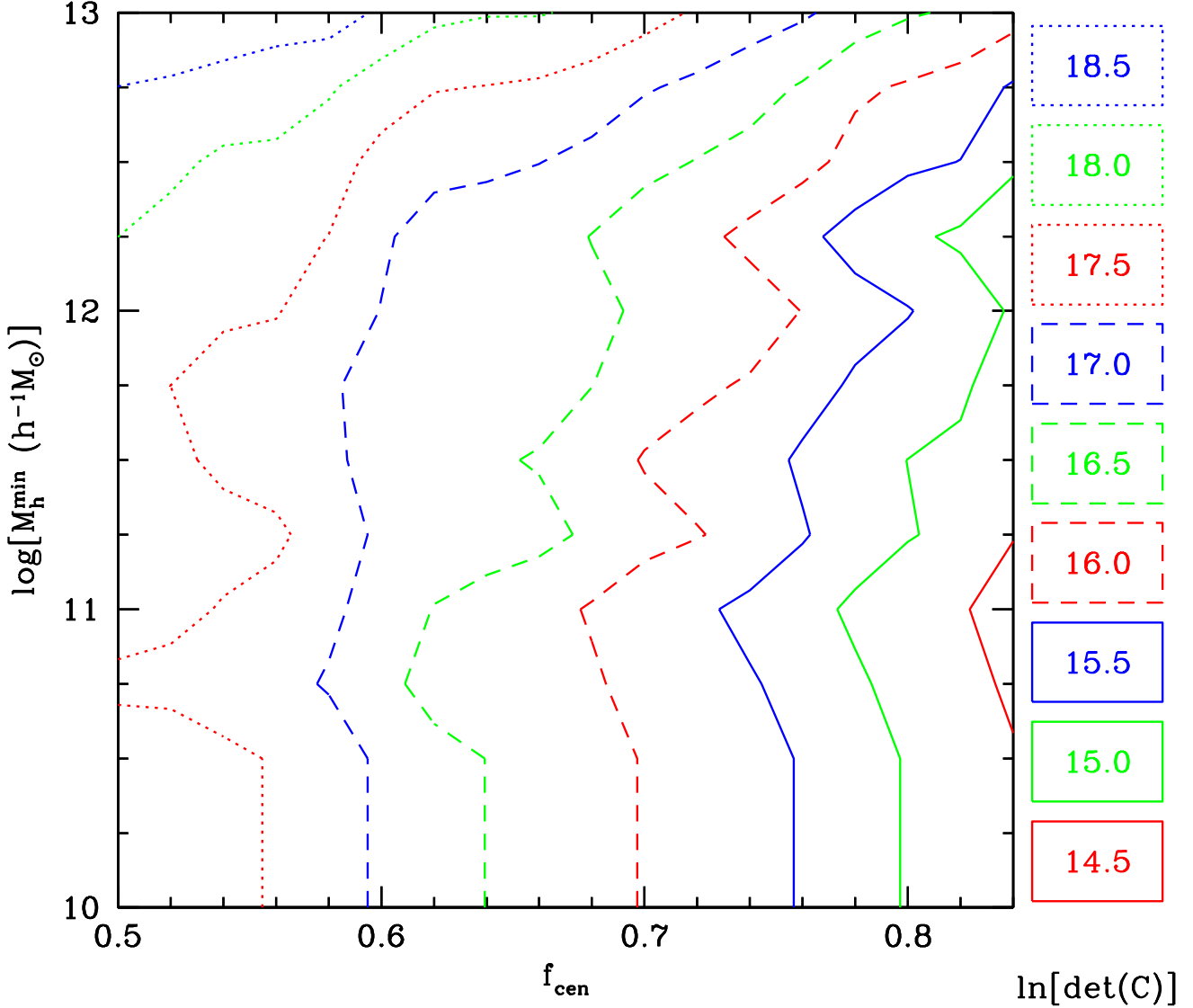,clip=true,width=\textwidth}}
\caption{Determinant of the covariance matrix of $w_p(r_p)$ (see Eq. \ref{eqn:cov_matrix}),
on the grid  of   the  two  model   parameters,  $f_{cen}$  and  $M_h^{min}$.
$f_{cen}$ is the fraction of radio AGN  that are hosted by the central
galaxy of  their own dark matter halo,  and $M_h^{min}$ is the minimum
mass  for the halos  that can host  radio AGN. The  contour levels are
indicated  at the right-hand side.
}
\label{fig:det_contour_whole}
\end{figure*}

\begin{figure*}
\centerline{\psfig{figure=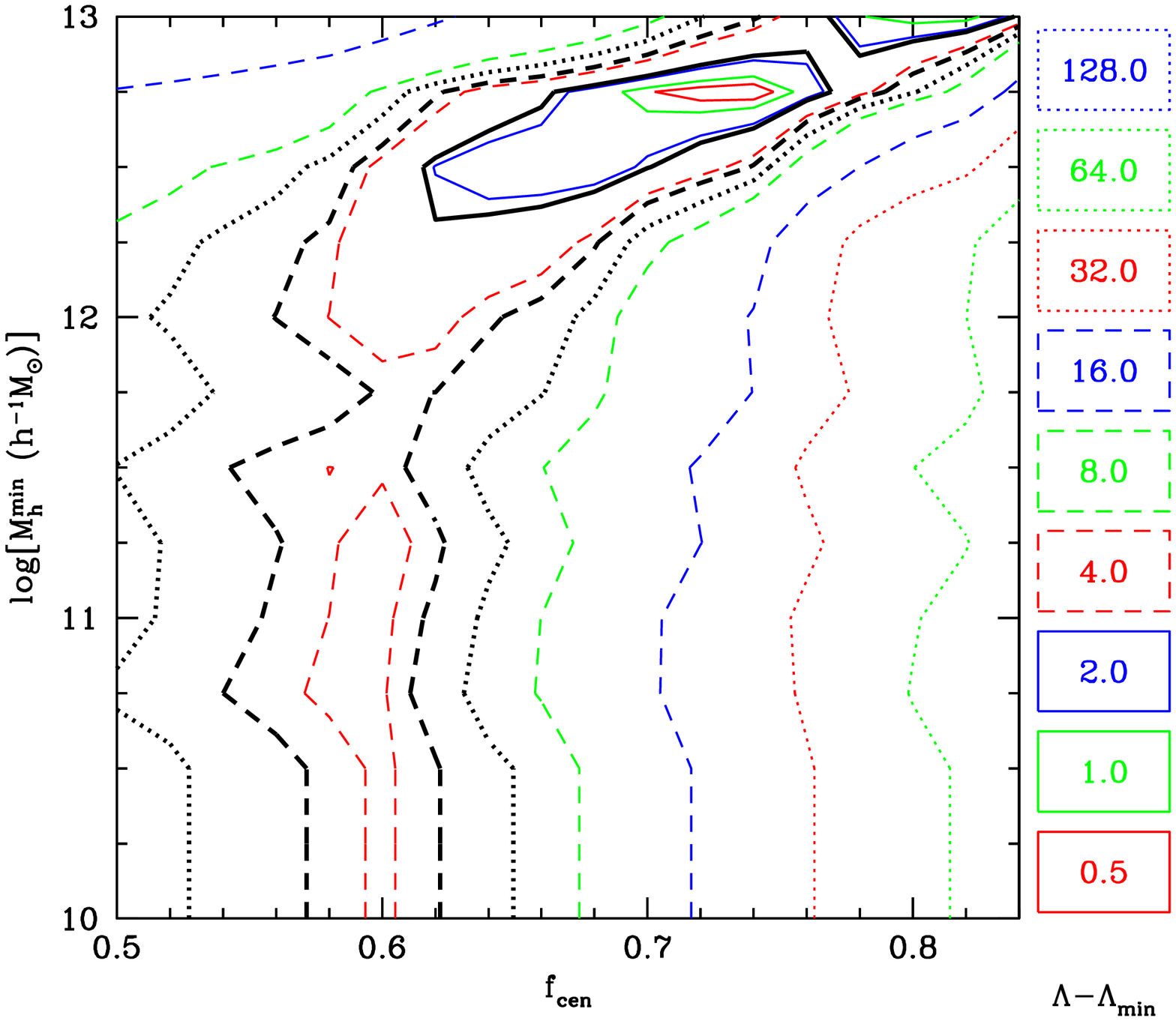,clip=true,width=\textwidth}}
\caption{$\Lambda$, defined by Eq.(\ref{eqn:lambda}) and derived by comparing 
the projected 2PCCF $w_p(r_p)$ for the whole sample of radio  AGN as predicted by the halo
occupation model and as measured from the SDSS data, is plotted with 
respect to the minimum value $\Lambda_{min}$, in the
grid  of   the  two  model   parameters,  $f_{cen}$  and
$M_h^{min}$. 
The  contour levels, as
indicated  at the right-hand side, are  chosen to cover the full range
of $\Lambda$ produced  by all  the  models.  The  68.3\%, 90\% and 95.4\%
confidence levels are plotted as solid, dashed and dotted black lines.
}
\label{fig:lambda_contour_whole}
\end{figure*}

We have measured the projected 2PCCF $w_p(r_p)$  of the radio AGN with
respect to the reference galaxies,  and compared  this to the  average
result  of the   five  control samples.  The  results    are  shown in
Fig.~\ref{fig:wrp_model_B} for the whole  sample (circles in  the
 left  panel)  and  for  two  different  ranges  in stellar mass
$M_\ast$.

Radio AGN are more strongly   clustered than control galaxies on   all
scales.  The difference in  $w_p(r_p)$ amplitude is a constant factor on scales
smaller  than $\sim$ 1  Mpc, then rises  slightly on larger scales. 
Fig.~\ref{fig:wrp_model_B}   also   shows    that the  clustering
amplitude of radio  AGN increases with   the stellar mass of  the host
galaxy.  This result is consistent with the fact  that more massive galaxies
are  more  strongly clustered  \citep{Li-06a}.    We have also  tested
whether there is a dependence of the clustering amplitude on the radio
luminosity  of the AGN, and we  find that at   fixed $M_*$, there is no
significant  effect.  This finding is  consistent with the   recent results of
\citet{Kauffmann-Heckman-Best-08}, who show that radio-loud AGN are in
denser environments than control radio-quiet  galaxies, but that there
is no dependence of local density on the radio luminosity of the AGN.

We  now use our  mock catalogues   (see  \S\ref{S:data}) to model  the
observed clustering measurements of radio AGN.  We first tried to vary
the fraction   of radio AGN   assigned   to central  versus  satellite
galaxies  (as was done  for the optical  AGN).  We found that if radio
AGN are preferentially found in  satellite galaxies, we can fit the
data   on  scales smaller    than a  few    Mpc,  but  the model  then
underpredicts the clustering amplitude  on larger scales.  As we have
discussed, the amplitude of the correlation  function on scales larger
than  a  few Mpc provides   a direct measure of   the mass of the dark
matter  halos  hosting the radio   AGN. Motivated  by  the  models  of
\citet{Croton-06} and \citet{Bower-06}, we impose a lower threshold in
halo mass, $M_h^{min}$, as a  second free parameter of  the model.  In
other words, radio-loud  AGN are only  found in dark matter halos more
massive   than  $M_h^{min}$.  The  probability  of a   galaxy to  be a
radio-loud AGN   depends not only   on  whether  it is  a  central  or
satellite system, but also on the mass of its dark matter halo.
However, the probability that a particular galaxy is an AGN does not
depend on the AGN status of its neighbors.  While
this step function in mass  is undoubtedly an over-simplification,  we
adopt it as a first attempt at modeling  to see if  it is close enough
to reality that the observations can be modeled in this way.

We  have  generated a  grid  of 322  models by 
varying the two parameters,  $f_{cen}$ and $M_h^{min}$, with $f_{cen}$
ranging  from  0.40   to  0.84  with  a  step  size of 0.02, and
$\log(M_h^{min}/h^{-1}M_\odot)$ ranging from 10.0 to 13.25 with a step size of
0.25.  We have constructed 200 mock catalogues of radio
AGN for each of the models.
We measure $w_p(r_p)$ and its covariance matrix $\mathbf{C}$ at each grid point,
and we compare the measurements to the SDSS results.
In order to identify the best-fit model, we first calculate the likelihood
of each parameter set, $L(f_{cen},M_h^{min})$, by
\begin{equation}
L(f_{cen},M_h^{min}) = \frac{\left(2\pi\right)^{-m/2}}{\det[\mathbf{C}(f_{cen},M_h^{min})]}
e^{-0.5\chi^2(f_{cen},M_h^{min})},
\end{equation}
where
\begin{equation}
  \chi^2(f_{cen}, M_h^{min})= 
  \mathbf{X}^T\mathbf{C}^{-1}\mathbf{X}.
\end{equation}
Here $\mathbf{X}=\{X_{j}\}\ (j=1,...,m)$ is an $m\times 1$ vector
with
\begin{equation}
  X_{j} = \left(\frac{1}{n}\sum_{i=1}^{n} w_{p,i}(r_{p,j})_{model}\right) - w_p(r_{p,j})_{SDSS},
\end{equation}
where $n=200$ is the number of mock catalogues, $m$ is the number of
radial bins over which $w_p(r_p)$ is measured, $w_p(r_{p,j})_{SDSS}$ is the 
clustering amplitude at the $j^{th}$ radial bin as measured from the SDSS,
and $w_{p,i}(r_{p,j})_{model}$ is the result at the $j^{th}$ radial bin
as measured with the $i^{th}$ mock catalogue.
The $m\times m$ matrix $\mathbf{C}=\{C_{ij}\}\ (i,j=1,...,m)$ is the
covariance matrix of the measurements from the 200 mock catalogues, given by
\begin{equation}\label{eqn:cov_matrix}
  C_{i,j} = \frac{1}{n-1}\left[\sum_{k=1}^n\left( Y_{ki}-\langle
      Y_i\rangle \right)\left(Y_{kj}-\langle Y_j\rangle \right)\right]
\end{equation}
where 
\begin{equation}
  Y_{k,i} = w_{p,k}(r_{p,i})_{model}
\end{equation}
is the measurement at the $i^{th}$ radial bin from the $k^{th}$ mock
catalogue, and 
\begin{equation}
\langle Y_i \rangle = \frac{1}{n}\sum_{k=1}^{n} w_{p,k}(r_{p,i})_{model}
\end{equation}
is the mean measurement at the $i^{th}$ radial bin over all mock
catalogues.  

We define the best-fit model  to be the  one
giving a minimum $\Lambda$ computed as follows:
\begin{eqnarray}\label{eqn:lambda}
&&\Lambda(f_{cen},M_h^{min}) = -2\ln L(f_{cen},M_h^{min}) \\
&&= \chi^2(f_{cen},M_h^{min})+2\ln\{\det[\mathbf{C}(f_{cen},M_h^{min})]\}+m\ln(2\pi).\nonumber
\end{eqnarray}
Note that this maximum likelihood estimate differs from the simple
minimum of $\chi^2$ if the determinant of the covariance matrix
$\det[\mathbf{C}]$ varies with the parameters.  This variation is
demonstrated in Figure \ref{fig:det_contour_whole}, where we plot
$\ln[\det(\mathbf{C})]$ in the grid of the two model parameters,
$f_{cen}$ and $M_h^{min}$. As can be seen, the determinant of the
covariance matrix does vary systematically from model to model.  This
variation is due to the fact that the covariance depends on the two-
and four-point functions of the galaxy distributions, which clearly
differ across the grid due to the different ways the halos are
populated with AGN.  However, the variation is relatively smooth,
indicating that we have used enough mock catalogues at each grid point
(200) to determine the covariance matrix with a sufficiently small
noise level.

We compare the measured $w_p(r_p)$ with the models in four radial bins
centered at $r_p=$ 0.21, 0.65, 2.1 and 6.5 $h^{-1}$Mpc, with a step
size of $\Delta\log r_p = 0.5$ (larger than the radial bins shown in
the plots). The choice of 4 radial bins was motivated by tests showing
that the covariance matrices evaluated from the mock catalogues were
well-behaved in this case, whereas using a significantly larger number
of radial bins causes the covariance matrices (a) to be noisier, and
(b) to have peculiar patterns of correlations between bins suggestive
of edge effects (in particular, strong correlations between certain adjacent
pairs of radial bins that are not representative of the overall
pattern of correlations).  These particular radial bins were chosen to sample separate
parts of the HOD, namely the central 1-halo term, the satellite 1-halo term, the
   transition between the 1- and 2-halo terms, and the 2-halo term
   (respectively).  The minimum radius was chosen because the 
   data at smaller separations are rather noisy.

Fig.~\ref{fig:lambda_contour_whole} plots the contours of
$\Delta\Lambda=\Lambda-\Lambda_{min}$ in the grid of the two
parameters, when using the $w_p(r_p)$ measurements for the full
radio-loud AGN sample (left panel of Fig.~\ref{fig:wrp_model_B}).  The
$1,2,$ and $3 \sigma$ confidence regions, computed for $m=4$ and 2
parameters, are indicated using solid, dashed and dotted black
lines. We have explicitly checked the distribution of $\chi^2$ values
for the individual mock catalogs to ascertain that the Gaussian
approximation for the likelihood is valid, and found that the
cumulative distribution of $\chi^2$ matches the expected distribution
at extremely high confidence (using the Kolmogorov-Smirnov test) at
all points on the grid.
Consequently, the 
method we have used to determine confidence regions is valid.  The
minimum $\Lambda_{min}$, appears at $f_{cen}=0.74$ 
and $M_h^{min}=10^{12.75}h^{-1}M_\odot$ with $\chi^2/d.o.f.=0.5$,
indicating that the fit is 
acceptable.  

\begin{figure}
\centerline{\psfig{figure=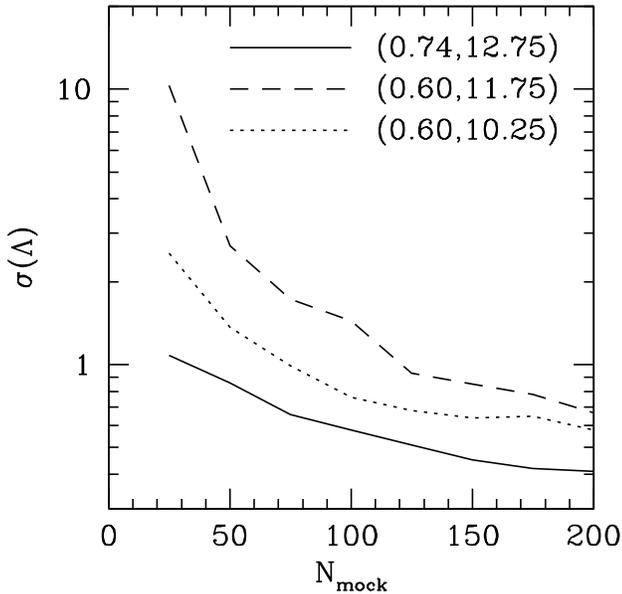,clip=true,width=\columnwidth}}
\caption{Noise in contour plot of $\Lambda$,
  Figure~\ref{fig:lambda_contour_whole}, due to finite $N_{mock}$, as
  determined from bootstrap 
  resampling of the 200 mock catalogs.  The noise as a function of
  $N_{mock}$ is shown for three different points on the grid: the
  global minimum, and two other points in the degeneracy region.}
\label{fig:sigma_contour_whole}
\end{figure}

However, there is a strong degeneracy between the two
parameters in the sense that models with smaller $f_{cen}$ and lower
$M_h^{min}$ can also provide a reasonable fit to the data.  
While this minimum is the preferred solution at the $1$-$\sigma$
level, there is a banana-shaped degeneracy region extending down to
$(0.60, 10.0)$ that is allowed at the $2$-$\sigma$ level.  The slight
saddlepoint at $(0.60, 11.75)$ that appears to divide this region does
not have a sufficiently large $\Delta\Lambda$ relative to the minimum
that we can robustly consider it is being real.  As shown in
Fig.~\ref{fig:sigma_contour_whole}, the residual noise in $\Lambda$
due to the use of finite $N_{mock}$ for the modeling is comparable to
the size of $\Delta\Lambda$ that creates this apparent saddlepoint.
The noise in $\Lambda$ was determined by bootstrapping the 
(roughly independent) $N_{mock}$ to make many new sets of mock
catalogs and determining the variance between $\Lambda$ for these
sets of $N_{mock}$. 
However, the size of $\Delta\Lambda$ that distinguishes this
degeneracy region from the rest of the plane in $(f_{cen}, M_h^{min})$
is significantly larger than the noise in $\Lambda$.  
The reason that we are able to fit the data reasonably
well at other points in this degeneracy region, despite the very
different halo model parameters, is that the 
higher satellite fraction places more AGN as satellites in very
massive halos, offsetting the lower bias for central AGN due to the
lower $M_h^{min}$.

We now use the center and right panels of Fig.~\ref{fig:wrp_model_B}
to evaluate whether the model that best fits the data for the full
sample can also describe the data for the sample split into two
stellar mass bins.  We have repeated the above analysis
with two samples of radio AGN with $\log M_\ast<11.44$ and $\log
M_\ast\ge 11.44$.  We evaluate the correlation function
using the same radial bins for 200 mock catalogues.  We see that while
the model is able to describe the data for the lower stellar mass bin
quite well, including the significant separation in signal between
radio AGN and controls, there is some tension between the observations
and the model for the higher
stellar mass bin.  As shown, the observed signals still have a significant
offset, but the model signals are nearly the same for the radio AGN and
for the controls.  This is not surprising, since at these high stellar
masses, essentially all halos are above $M_h^{min}$.  Consequently,
the apparent failure of the model at high stellar mass most likely
results from the fact that a step-function model for the radio AGN
probability is overly simplistic.  A probability that is a function of
mass would allow for a better description, but unfortunately the data
quality do not justify adding additional halo model parameters at this
time, so we defer such an analysis to future work with more data.

\subsubsection{Galaxy-galaxy lensing}\label{S:radioAGN_gglensing}
\begin{figure*}
\includegraphics[width=6in,angle=0]{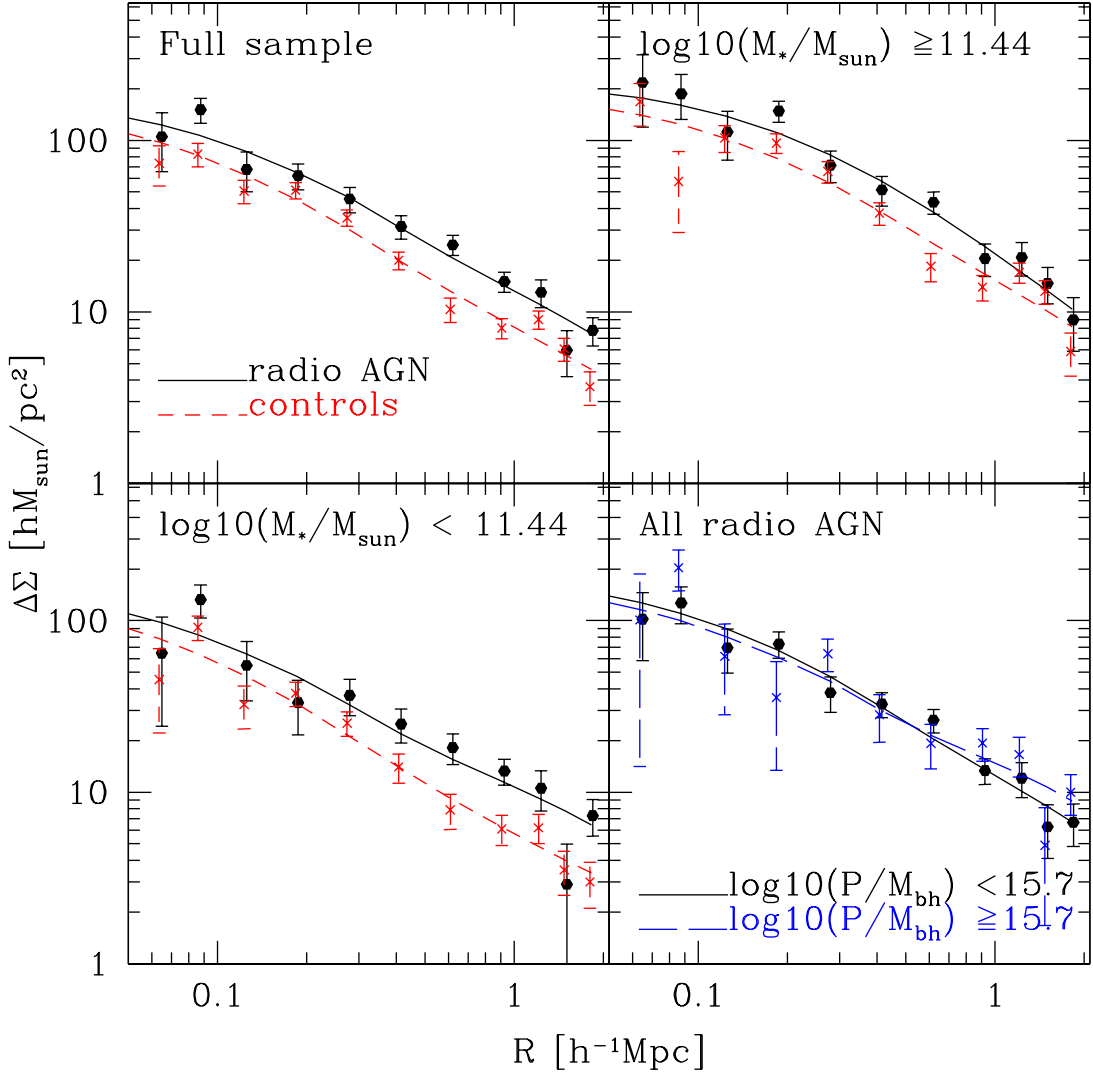}
\caption{\label{F:gg-rAGN}The galaxy-galaxy lensing signal for
  the radio-loud AGN and control galaxies split into subsamples as
  indicated on the plot.  Points show the measured signal and lines show the
  best-fitting halo model.} 
\end{figure*}

Here we present the galaxy-galaxy weak lensing signal for the radio-loud
AGN sample, along with the halo model fits.  These results are shown
in Fig.~\ref{F:gg-rAGN} for the full
radio-loud AGN and control samples (upper left); a stellar mass subsample containing
the lower $2/3$ of the sample in stellar mass (lower left); the remaining upper
$1/3$ of the sample in stellar mass (upper right); and the radio-loud
AGN split by
$\log{(P/M_{bh})}$ (lower right).  

\begin{figure*}
\includegraphics[width=\textwidth,angle=0]{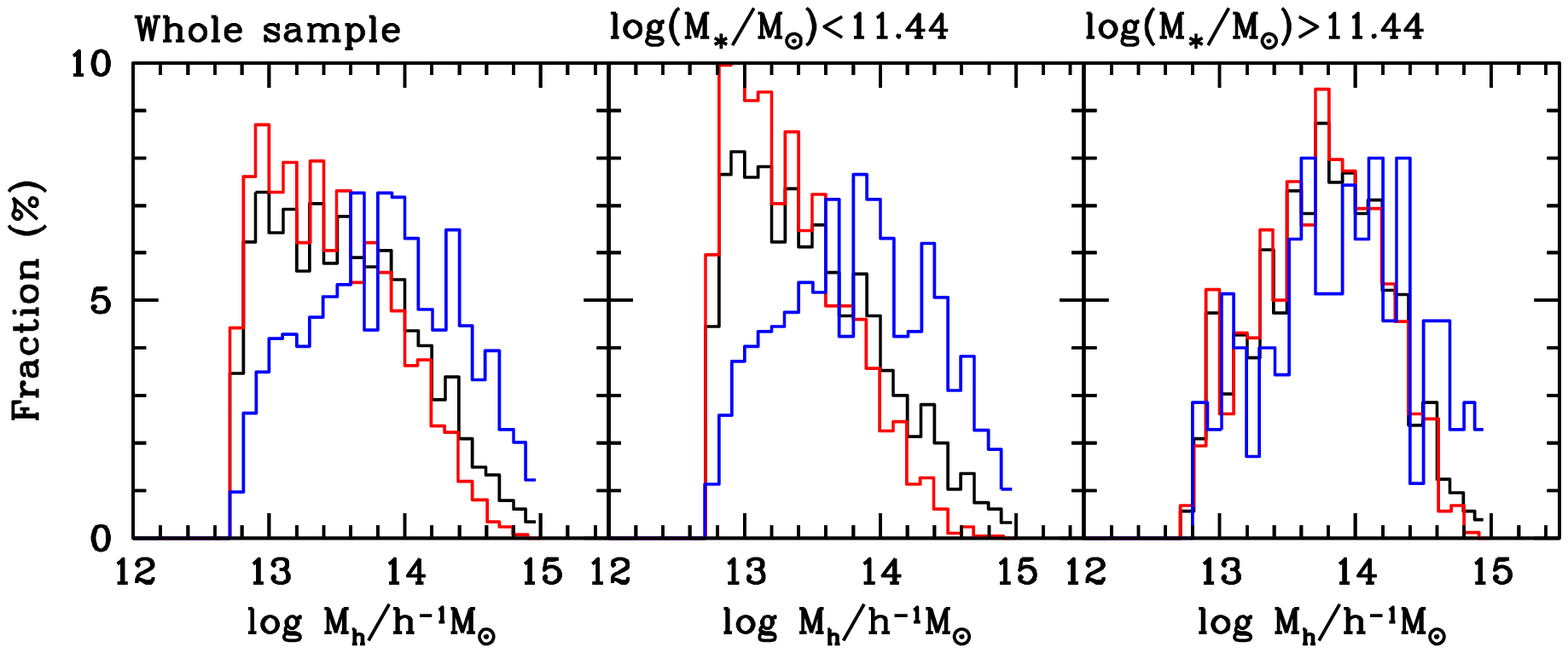}
\caption{\label{F:mcentdist-ragn}The distributions of halo
  mass in the mock catalogues that are able to reproduce the clustering
  signal for radio-loud AGN in the full sample and our two stellar
  mass bins.  Results are shown for the full sample (black lines),
  central galaxies only (red lines), and satellites only (blue
  lines).} 
\end{figure*}

A number of trends are evident in
Fig.~\ref{F:gg-rAGN}.  First, we see that the lensing signal is higher
on all scales ($R<2h^{-1}$Mpc) for the radio-loud AGN than for the control sample.
Second, this trend persists in the different stellar mass subsamples.  Third, the
lensing signal increases with stellar mass, as expected.  Finally, the
lensing signal for the two bins in $\log{(P/M_{bh})}$ are not markedly
different from each other.  For the lower and higher bins in this
quantity, the mean 
stellar masses are $2.45$ and $2.75\times 10^{11}M_{\odot}$,
respectively, a difference of only 10 per cent, much less than the
difference between the mean stellar masses for the two stellar mass
bins, so this near equivalence is not surprising. 

Next, we present the halo model interpretation of these results.
Table~\ref{T:radio-gg-hmfits} gives the best-fitting halo model parameters
for each of the subsamples shown in Fig.~\ref{F:gg-rAGN}.  
\begin{table*}
\caption{\label{T:radio-gg-hmfits}Best-fitting halo model parameters for
  fits to the radio-loud AGN g-g weak lensing signal, with 68 per cent
  CL errors (in each case, marginalized over the other HOD parameter).}
\begin{tabular}{lccccc}
 & \multicolumn{2}{c}{Radio-Loud AGN} & \multicolumn{2}{c}{Controls} & \\
Sample & $M_{cent}$ &  $\alpha$ & $M_{cent}$ &  $\alpha$ &
$p(M_{cent}^{(AGN)} > M_{cent}^{(control)})$ \\
 & $[10^{13} h^{-1}M_{\odot}]$ & & $[10^{13} h^{-1}M_{\odot}]$ & & \\
\hline
\hline
Full & $1.6\pm 0.4$ & $0.22\pm 0.11$ & $0.91^{+0.12}_{-0.10}$ &
$0.13\pm 0.05$ & $0.97$ \\
$\log{(M_*/M_\odot)} < 11.44$ & $0.8^{+0.4}_{-0.5}$ &
$0.31^{+0.30}_{-0.16}$ & $0.56^{+0.11}_{-0.09}$ &
$0.10^{+0.05}_{-0.07}$ & $0.71$ \\
$\log{(M_*/M_\odot)} \ge 11.44$ & $4.9^{+0.7}_{-0.9}$ &
$0.01^{+0.15}_{-0.01}$ & $2.5\pm 0.4$ & $0.16\pm 0.10$ & $0.99$ \\
$\log{(P/M_{bh})} < 15.7$ & $1.8\pm 0.4$ & $0.13^{+0.15}_{-0.13}$ & - & - & - \\
$\log{(P/M_{bh})} \ge 15.7$ & $1.4^{+0.4}_{-0.9}$ & $0.40^{+0.40}_{-0.15}$ & - & - & - \\
\hline
\hline
\end{tabular}
\end{table*}
We focus on the results for central halo masses, because the satellite
fractions are quite noisy.  For the full sample, the best-fitting
central halo mass $M_{cent}$ is 80 per cent higher for the radio-loud AGN
than for the control sample at fixed stellar mass and redshift.
When accounting for statistical correlations between the two samples
using the bootstrap method, we find that  the central radio-loud AGN
have higher mass than the controls at the 97 per cent
confidence level.  For the lower stellar mass bin, this conclusion is
less statistically significant (50 per cent higher mass, with
$p(M_{cent}^{(AGN)}>M_{cent}^{(controls)}) =0.71$) but for the higher
stellar mass bin, it is more significant than for the full sample
(factor of two higher mass, with
$p(M_{cent}^{(AGN)}>M_{cent}^{(controls)}) =0.99$).  This result is 
in sharp contrast to the results for central optical AGN, 
which appear to have the same halo mass as the optical control
galaxies.

When comparing the amplitude of the central halo masses to those
for the general galaxy population studied via lensing in
\cite{2006MNRAS.368..715M}, we find that the 
results for the control sample are consistent with that paper once we
account for the different halo mass definitions,
{\it provided} that we compare 
to the early type galaxy sample (which has higher mean central
halo mass than the late type sample for $M_*>10^{11}M_{\odot}$).  

Given the high mean halo mass inferred for central radio-loud AGN, we
must check our modeling assumption that the range of halo masses is
narrow.  Fig.~\ref{F:mcentdist-ragn} shows a plot of the halo mass
distribution for model B described in
section~\ref{S:radioAGN_gclustering} which is able to reproduce the
radio AGN clustering results.  As discussed, this HOD includes a
minimum halo mass cutoff of $10^{12.5}h^{-1}M_{\odot}$.  As can be
seen, the central halo mass distributions are indeed broader than the
factor of $\sim 6$ that we need for the $M_{cent}$ derived from
lensing to be physically meaningful.  As before, we do not apply a
correction factor, due to the uncertainty in determining it.  We
merely note that since the AGN are more likely than the controls to
reside in clusters (a claim that we will back up via direct comparison
with a cluster catalogue), the correction factor to get the mean
central halo mass should be {\em higher} for the AGN than for the
controls. Thus, the size of the central halo mass difference between
radio-loud AGN and control galaxies is in fact underestimated by our
neglect of these corrections.

To avoid uncertainties in the best-fitting halo masses caused by the
cluster membership of some galaxies, we have cross-correlated our
radio-loud AGN and control samples with a pre-existing SDSS sample of
galaxy clusters. We can then compare halo masses for radio-loud AGN
and control galaxies in the field (i.e. excluding cluster members).
For this sample, the distribution of halo masses will no longer have a
tail extending to very high masses.

Since the radio-loud AGN sample is predominantly ($85$ per cent) in
the redshift range $0.1<z<0.3$, the natural choice of cluster catalogue
is the SDSS MaxBCG catalogue \citep{2007ApJ...660..221K,2007ApJ...660..239K},
which contains clusters in the redshift range
$0.1<z<0.3$ that are  selected based  
on the existence of a red sequence. 

We select radio-loud AGN and control galaxies in this redshift range  
and check whether they are within
$1h^{-1}$Mpc (physical projected separation) of a cluster, and within
$\Delta z=\pm 0.04$ of the cluster BCG. We note that the choice of redshift
separation is a factor of ten larger than the velocity dispersion
of even the very largest galaxy clusters. The motivation
for this choice  comes from the typical
photometric redshift error of the galaxies in the maxBCG catalogue, 
and  ensures that 95 per cent of true cluster members  would be
found.  For reference, the minimum mass of the public maxBCG catalogue,
with scaled richness $\ge 10$, is $\sim 6 \times 10^{13}
h^{-1}M_{sun}$ (defined using $M_{200\overline{\rho}}$,
\citealt{2008arXiv0802.2365R}). 
We also determined whether the radio-loud AGN or control is the cluster
BCG or a satellite.  These statistics are presented for radio-loud AGN and
controls for both the full sample and for subsamples selected
according to stellar mass in
Table~\ref{T:maxbcgmatch}.  

\begin{table*}
\caption{\label{T:maxbcgmatch}Results of matching the radio-loud AGN and
  control samples with $0.1<z<0.3$ against the maxBCG cluster
  catalogue.  Quantities presented in the table are defined in
  Equations~\ref{E:deffclust} and~\ref{E:deffbcg}.}
\begin{tabular}{lcccc}
Sample & $\fclust$ (AGN) &  $\fclust$ (controls) & $\fbcg$ (AGN) &  $\fbcg$ (controls) \\
\hline
\hline
Full & $0.24$ & $0.16$ & $0.14$ & $0.09$ \\
$\log{M_*} < 11.44$ & $0.20$ & $0.12$ & $0.08$ & $0.05$ \\
$\log{M_*} \ge 11.44$ & $0.36$ & $0.29$ & $0.31$ & $0.23$ \\
\hline
\hline
\end{tabular}
\end{table*}
The numbers presented there for each sample are defined as follows:
\begin{equation}\label{E:deffclust}
\fclust = \frac{\mbox{Number in sample that are in a
    cluster}}{\mbox{Number in sample}}
\end{equation}
and
\begin{equation}\label{E:deffbcg}
\fbcg = \frac{\mbox{Number in sample that are
    BCG of a cluster}}{\mbox{Number in sample}}
\end{equation}

Based on this table, we note a few interesting trends.  First, for all
subsamples we considered, $\fclust$ and $\fbcg$
are higher for the radio-loud AGN than for the controls.  This difference
is most pronounced for the lower stellar mass bin, with a
factor of two difference between radio-loud AGN and controls.  For the
higher stellar mass bin, the differences are at the $\sim 20$ per cent
level. These results are broadly in agreement with those of
\cite{2007MNRAS.379..894B}.    

Second, if we compare these two numbers for a given sample, we
can determine the fraction of those {\it in clusters} that are BCGs,
where the rest are satellites.  For the full radio-loud AGN and
the control samples, this 
number is 60  per cent.   Thus, on average, the distribution of
centrals versus satellites for those that are cluster members is the
same for radio-loud AGN and control galaxies.  For the lower and higher stellar mass samples,
we again find consistency between the radio-loud AGN and controls, with BCG
fractions of those that are in clusters of 40 per cent and 85 per
cent, respectively.

It is apparent that the halo mass distributions of both the radio-loud AGN
and control samples may have significant contributions from cluster
BCG and satellite galaxies, which will skew the halo mass distribution
to the high mass end.  Consequently, we repeat  the weak lensing analysis
using only those galaxies in the redshift range
$0.1<z<0.3$ that are not within any maxBCG
cluster.  This cuts down the size of the sample significantly at the
high stellar mass end, so we
only analyze the full sample,
without any divisions in stellar mass or radio power.

\begin{figure}
\includegraphics[width=3.3in,angle=0]{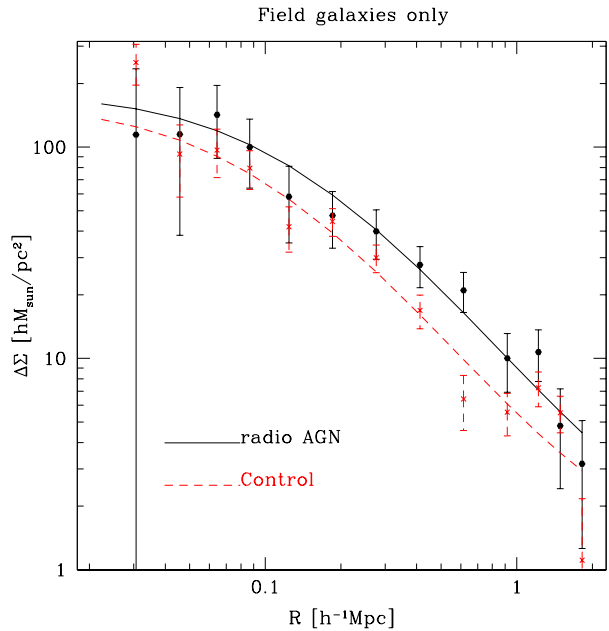}
\caption{\label{F:gg-rAGNfield}The galaxy-galaxy lensing signal for
  the radio-loud AGN and control galaxies located in the field with
  $0.1<z<0.3$.  Points show the measured signal and lines show the
  best-fitting halo model.} 
\end{figure}

The lensing signal for these ``field samples'' is shown in
Fig.~\ref{F:gg-rAGNfield}, along with the 
best-fitting halo model.  We initially used the same halo model as before,
but found that the satellite fractions
were all consistent with zero within the noise (as one would expect
given the sample design).  Thus, we redid the fits with fixed
$\alpha=0$, fitting for central halo mass only.

The mean stellar mass for these field samples is 
$2.1\times 10^{11}M_{\odot}$.  The reduction in best-fitting
central halo mass going from the full radio-loud AGN and control samples,
to the field subsamples, is roughly 15 per cent.  This reduction is
expected, since we have excluded the AGN  in clusters, which have
typical halo masses of $> 10^{14} M_{\odot}$.  In both cases,
however, for the full samples and for the field subsamples, the best-fitting
central halo mass for the radio-loud AGN sample is roughly twice that of
the control samples.  We have already explored this result for the
full sample; for the field sample, we find best-fitting central halo
masses of $(1.5\pm 0.3)$ and 
$(0.76\pm 0.14)\times 10^{13} h^{-1}M_{\odot}$ (radio AGN and controls,
respectively). The
difference between the two masses is 
thus significant at the 97 per cent CL. 

\subsection{Joint constraints on halo masses and satellite fractions of radio AGN}
\label{S:joint_constraints}

\begin{figure}
\includegraphics[width=\columnwidth,angle=0]{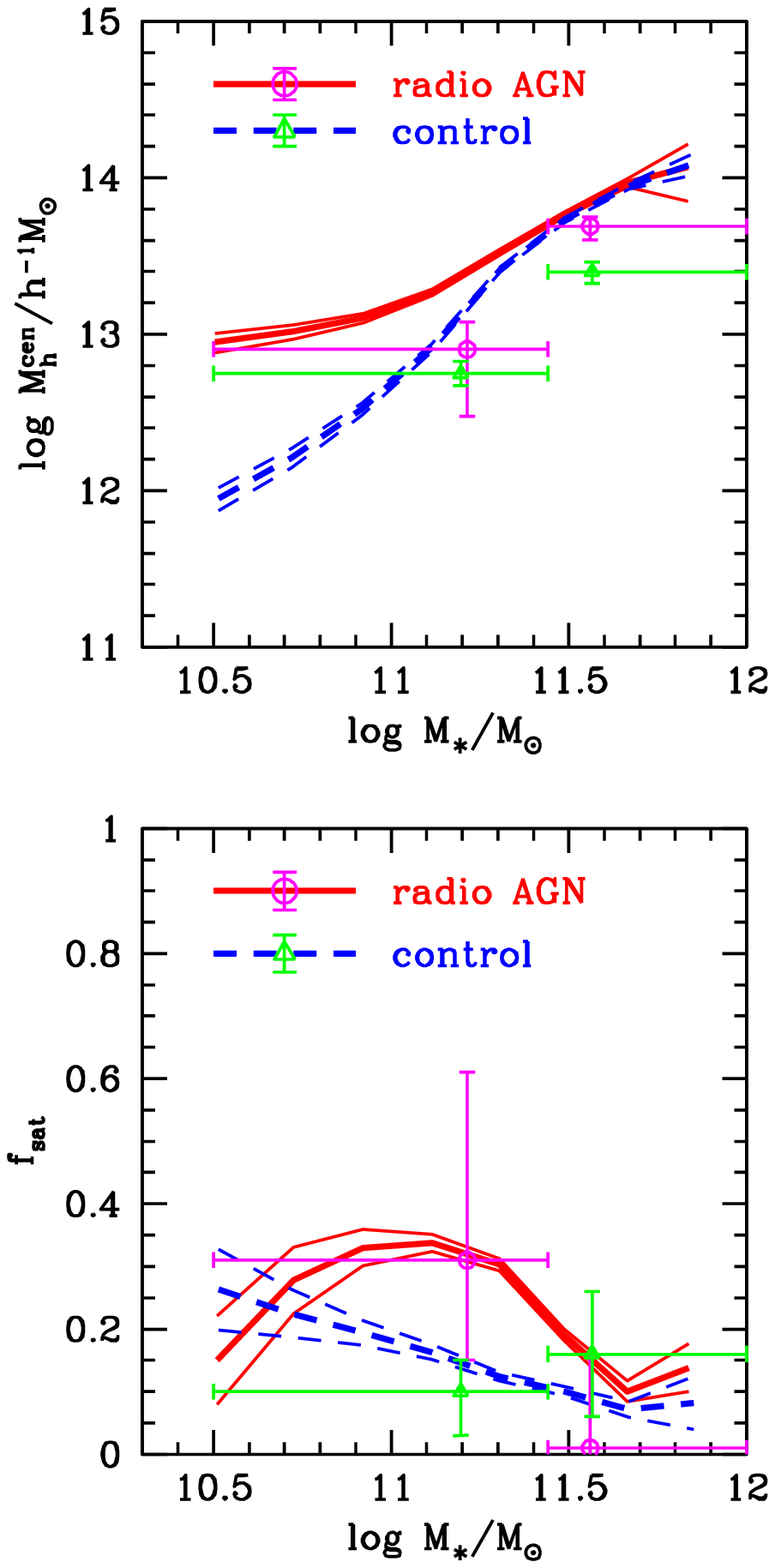}
\caption{\label{fig:model_pars_smass}
Mean halo  mass for central  radio  AGN  (upper panels) and  satellite
fraction of  all radio AGN (lower panels)  are plotted  as function of
stellar mass.   In  each   panel,  the  red    thick  line shows   the
best-fitting model determined by the  clustering measurements of radio
AGN,   and the   blue thick   line    shows the   result  for  control
galaxies. The thin lines show the $1-\sigma$  variance between 200 mock
catalogues.  The  results determined  by the g-g  lensing
analyses are plotted  as magenta circles  for  radio AGN and as  green
triangles for  control  galaxies;  horizontal errorbars indicate   the
widths of the stellar mass bins.}
\end{figure}

In  section~\ref{S:radioAGN_gclustering},   we  presented  a
best-fitting HOD model that  well reproduced the clustering
signal of  radio AGN  and  the control sample.   In  this  section, we
compare  the results  of the model  to  the results obtained from  the
lensing  analysis. 

In Fig.~\ref{fig:model_pars_smass}, we plot the mean halo mass of 
the central radio-loud AGN and the fraction of satellite AGN, as predicted 
by the best-fitting model describing the clustering results, as a
function of  stellar mass.  
The results from the lensing analyses are also plotted.  
 As discussed in the previous section,
the lensing analysis finds a higher mean halo mass for radio AGN
at slightly more than $2\sigma$, 
which persists even when radio AGN residing in clusters
are excluded from the analysis.  This difference is even more
significant, at the 99 per cent CL, for the higher mass subsample.

As shown, the model that best describes the clustering results is
consistent with the g-g lensing results in the sense that different
mean masses are predicted for the controls and the radio AGN.  As
already discussed, the fact that the same masses are predicted for the
two samples in the higher mass bin is a consequence of
overly-simplistic modeling.  While the g-g lensing estimates of the
satellite fraction are fairly noisy, they are also consistent with the
model that describes the clustering.  It would be valuable to confirm our conclusions using larger 
radio AGN samples that will be available in the future.  This is
particularly true at very low stellar masses ($<10^{11}M_{\odot}$),
where a measurement with 
lensing was not at all possible and where we would be most sensitive
to the difference between radio AGN and controls.  When samples are
available with significantly 
better statistics, the data quality will warrant a more careful
analysis using the mocks to compare against both the clustering and
lensing signal, and possibly more sophisticated halo modeling than
that which was attempted here.

\section{Summary}\label{S:summary}

\begin{figure*}
\centerline{
\psfig{figure=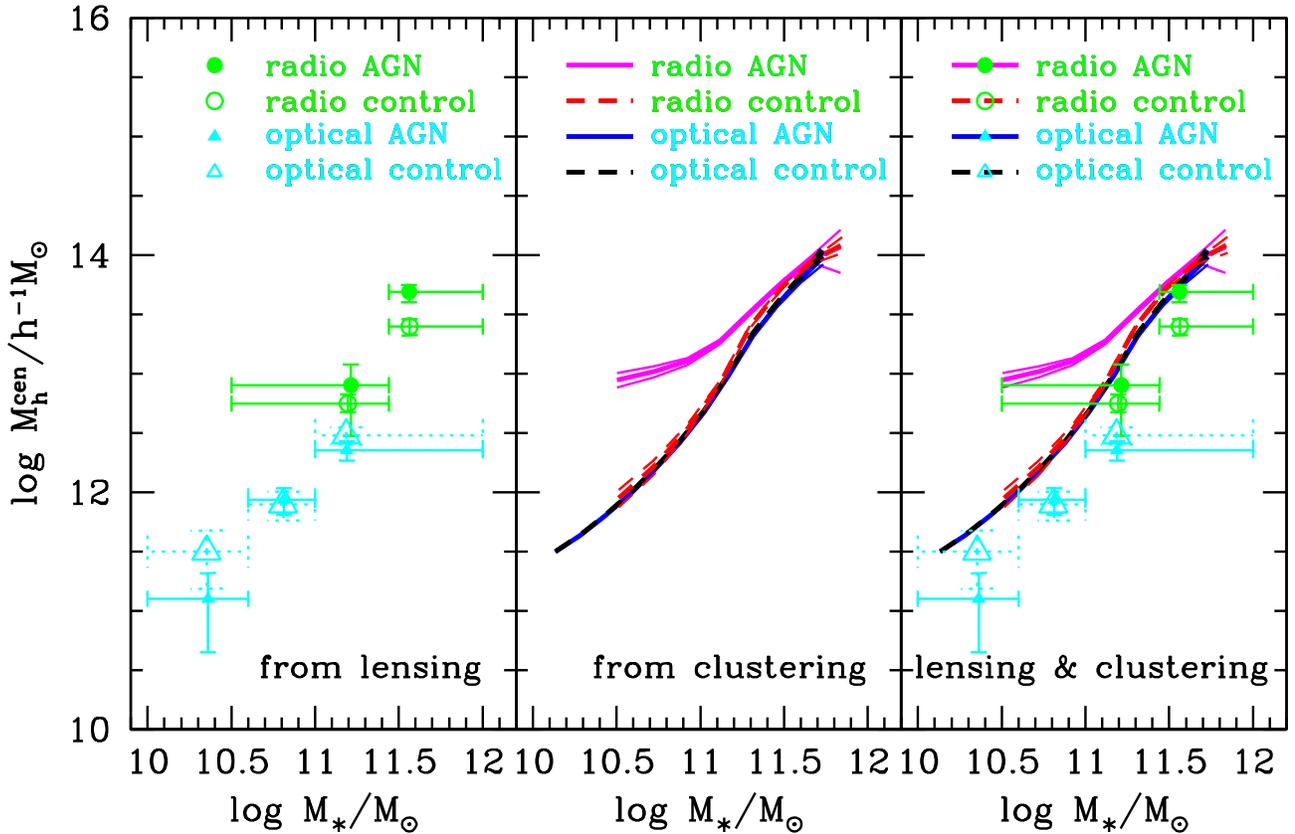,clip=true,width=\textwidth}
}
\caption{{\em Left}: Average central halo mass derived from galaxy-galaxy lensing 
is plotted as function of stellar mass, for three subsamples of optical AGN 
(cyan) and two subsamples of radio AGN (green). Filled symbols  show results
for the AGN samples, whole open symbols show results for the control
samples.  The stellar mass bin widths are indicated with horizontal
errorbars. 
{\em Middle}: Average central halo mass derived from the clustering analysis 
is plotted as a function of stellar mass  for optical (blue) and
radio (magenta) AGN, and for the control galaxies
(black dashed line for optical and red dashed for radio).  
{\em Right}: The results from the lensing and the clustering analyses are
plotted on top of each other for comparison.
}
\label{fig:avg_hmass}
\end{figure*}

We now present a comparison of the inferred  halo masses and satellite
fractions of both optical and radio-loud AGN, from both the clustering
and g-g lensing techniques.

In the left panel of 
Fig.~\ref{fig:avg_hmass}, we plot the inferred central halo masses
of optical AGN (cyan) and radio AGN (green) from the lensing analysis as
a function of stellar mass. The solid symbols give results for the
AGN, while the open symbols give results for the control samples.

In the middle panel, we plot the inferred central halo masses of
optical (blue) and radio AGN (red) from the clustering analysis. The
results are from the published model of \citet{Li-06c} for the optical
AGN, and the magenta lines show our best-fitting model for
the radio AGN presented in section~\ref{S:radioAGN_gclustering}.  The
halo mass as a function of stellar mass is plotted as solid lines for
the AGN and as dashed lines for the control galaxies. As can be seen,
the clustering technique allows us to extend the halo mass
measurements over a significantly larger range in stellar masses.  We
have experimented with subsamples with even lower stellar masses and
we found that we are able to derive a (noisy) clustering measurement
for radio galaxies with stellar masses less than $10^{11} M_{\odot}$,
but there are too few objects to permit a lensing analysis to be
carried out.  Lensing measurements at lower stellar mass with larger,
future datasets may be critical for reducing the modeling uncertainty
indicated in this figure. Note that the convergence of the curves at
high stellar mass is a consequence of our very simple model which has
little or no flexibility to adjust the clustering of high mass stellar
mass objects, the great majority of which are central galaxies and
live in massive halos.

Finally, in the right-hand panel, we superimpose the clustering and lensing
results. We see that the agreement between the results for these
two completely independent techniques is satisfactory, once we take
into account the fact that (a) there is some modeling uncertainty in the
best-fitting central halo masses from lensing due to the assumption of
a narrow central halo mass distribution, and (b) the halo masses from
the lensing analysis are relatively independent of the assumed
$\sigma_8$ (which to lowest order only affects the best-fit $\alpha$),
but the halo masses from the clustering analysis are tied to
$\sigma_8=0.9$ from the Millennium simulation through the large-scale
bias - halo mass connection.  In the latter case, we can estimate the
effect of lowering $\sigma_8$ from $0.9$, as in the mocks, to $0.8$
(as in the WMAP 5-year results, \citealt{2008arXiv0803.0586D}) at fixed $\Omega_m=0.25$.
Using the mass function  from \cite{2008arXiv0803.2706T} 
at the typical redshift of radio AGN, and translating to
our halo mass definition, we find that the requirement that the
stellar mass function be matched, which is essentially an abundance
constraint, would lead to the masses from the clustering analyses be
lowered by $\sim 15$ per cent at $10^{13}h^{-1}M_{\odot}$, or $\sim
25$ per cent at $3\times 10^{13}h^{-1}M_{\odot}$.  While this
difference is significant, it can only partially account for the
differences shown in the figures.  However, the modeling uncertainty due to
the simple HOD used for the lensing analysis can lead to significant
additional uncertainty in those masses, typically leading to
underestimation (i.e., the sign of the apparent discrepancy) by
several tens
of per cent when the central halo mass distributions are
quite broad \citep{2005MNRAS.362.1451M}, as is the case for several of
our stellar mass subsamples (Figs.~\ref{F:mcentdist-optagn} and~\ref{F:mcentdist-ragn}).  

Putting the lensing and the clustering results together leads us to
the following major conclusions:
\begin {itemize}
\item Radio AGN are hosted by galaxies with higher stellar masses than optical
AGN, and are also in more massive dark matter halos. The mean stellar
mass of the optical AGN sample is $8\times 10^{10}M_{\odot}$ and the
corresponding mean central  
halo mass deduced from galaxy-galaxy lensing is $(8.0\pm 1.5)\times
10^{11}h^{-1}M_{\odot}$. 
The mean stellar mass of the radio AGN sample is $2.5\times
10^{11}M_{\odot}$ and the mean 
halo mass deduced from galaxy-galaxy lensing is $(1.6\pm 0.4)\times
10^{13}h^{-1}M_{\odot}$.  Thus, the mean stellar mass and mean central
halo mass for the radio AGN are $\sim 3$ and $\sim 20$ times the
corresponding values for the optical AGN.  

\item At {\em fixed} stellar mass, radio-loud AGN inhabit more massive dark matter
halos than optical AGN. This is seen both in the clustering and
and in the galaxy-galaxy lensing analyses. 
Note that for the g-g lensing analysis, the highest stellar mass 
bin for the optical AGN and the lowest bin for the radio-loud
AGN have the same mean stellar mass.  Fig.~\ref{fig:avg_hmass} shows
that these two samples have very different halo masses.  While the
halo masses are also different for the control galaxies due to the
morphology-dependence of halo mass at fixed stellar mass
$>10^{11}M_{\odot}$ \citep{2006MNRAS.368..715M}, the difference is
even more pronounced for the optical and radio-loud AGN samples.

\item At {\em fixed} stellar mass, optical AGN inhabit dark matter
  halos of similar mass as galaxies of the same stellar mass selected
  without regard to AGN properties.

\item At {\em fixed} stellar mass, radio-loud AGN inhabit more massive
dark matter halos than galaxies of the same stellar mass selected
without regard to AGN properties.  We emphasize that despite the
difficulty in representing this offset using our simple clustering
model, it is present observationally at high significance in both the
clustering data (Fig.~\ref{fig:wrp_model_B}) and the lensing data
(Figs.~\ref{F:gg-rAGN} and~\ref{F:gg-rAGNfield}) both for the full
samples and for the two stellar mass subsamples.

\item The clustering and lensing analyses together favour a model in which radio-loud AGN are not
found in dark matter halos with masses less than about $3 \times 10^{12} h^{-1}
M_{\odot}$, though the preference for this over a model with a minimum
mass a factor of $10$ smaller is only a $2\sigma$ difference given the
size of current datasets.

\end{itemize}

One unresolved point in the reconciliation between the modeling of the
clustering and its comparison with the lensing analysis is that the
lensing analysis found the most significant difference between the central
halo masses for radio AGN and controls at the high stellar mass end,
whereas the clustering modeling suggests the largest difference should
occur for lower stellar mass.  This point is simply an artifact of
overly simplistic modeling of a strict mass threshold, and may
therefore be resolved with more 
sophisticated modeling involving a probability that is a function of
mass; however, the introduction of more model 
parameters is not justified by the achievable $S/N$ of the data at
this time.

\begin{figure}
\centerline{
\psfig{figure=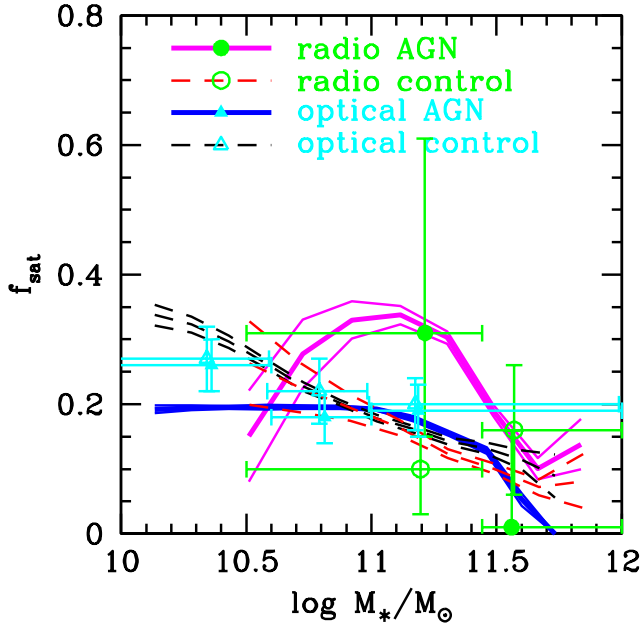,clip=true,width=0.5\textwidth}}
\caption{Satellite fraction as a function of stellar mass. Symbols and lines are the same as in the previous figure.}
\label{fig:avg_fsat}
\end{figure}

In  Fig. ~\ref{fig:avg_fsat}, we compare the satellite fractions 
inferred by the two techniques.
The results are broadly consistent with each
other. The clustering analysis yields bigger differences in the
inferred satellite fractions, particularly for AGN with low
stellar masses. Optical  AGN with low stellar masses are predicted
to be quite strongly biased to  central galaxy hosts,
whereas radio AGN with low stellar masses are predicted to be
located more frequently in satellite galaxies. 
The lensing analysis gives some weak
indications of trends in the same direction, but the results are far
from conclusive.

\section{Implications of this work}\label{S:implications}

Perhaps the most important finding of this work is that optical
AGN largely follow the same relation between stellar mass and halo mass
as ``ordinary'' galaxies, but that radio-loud AGN deviate significantly
from it (by a factor of $\sim 2$). This statement is true at the 97 per
cent CL even if we restrict the analysis to those radio-loud AGN and
control galaxies in the field (excluding those in massive groups and
clusters).                    

This result implies that the large-scale halo environment plays an
important role in understanding the radio AGN phenomenon. Previous
work \citep[e.g.][]{Best-05b} has shown that the fraction of radio AGN
increases strongly for more massive galaxies, suggesting that radio
jets are more readily triggered in galaxies with more massive black
holes.  However, when we compare radio AGN with control galaxies of
the same stellar mass selected without regard for nuclear activity, we
find that the radio AGN reside in dark matter halos that are a factor
of two more massive on average. This boost in halo mass appears to be
{\em largely independent of luminosity of the radio source.} This
demonstrates that black hole mass is not the only parameter that
controls the radio AGN phenomenon -- some aspect of the larger-scale
environment of the galaxy must play a crucial role in regulating when
the jet is switched on or when it is visible at radio wavelengths.

Recent semi-analytic models have assumed that feedback from radio AGN  
only becomes important in halos in which gas is cooling quasi-statically,
i.e. halos  above a mass of a few $\times 10^{11} -10^{12} M_{\odot}$
\citep{Croton-06, Bower-06, Cattaneo-06}.
Our clustering results strongly support this idea.
Our two best-fitting HOD models both invoke a minimum halo mass 
close to these values, below which
radio AGN are no longer found. If we do not impose a minimum mass,
our models are not  able to fit the observation correlation amplitude
of radio-loud AGN on large scales.

Finally, the fact that  the optical AGN in  our sample follow the same
$M_\ast-M_{halo}$ relation  as the  general galaxy population, implies
that  the optical AGN phenomenon  is  largely decoupled  from the host
halo.  For the highest stellar mass bin, the best-fitting central halo
mass  derived from  galaxy-galaxy   lensing is  more  consistent  with
published results for  late-type galaxies.  This result  is consistent
with the  general tendency of  narrow-line AGN  to be associated  with
galaxies  with ongoing  star   formation and  by  extension,   a  cold
interstellar medium.

\section*{Acknowledgments}
R.M.  is supported by NASA through Hubble Fellowship grant
\#HST-HF-01199.02-A awarded by  the Space Telescope Science Institute,
which is operated  by the Association of Universities  for Research in
Astronomy,  Inc.,  for NASA,  under  contract  NAS  5-26555.  C.L.  is
supported  by  the   Joint  Postdoctoral  Programme  in  Astrophysical
Cosmology  of  Max  Planck  Institute for  Astrophysics  and  Shanghai
Astronomical Observatory,  and by NSFC  (10533030, 10643005, 10633020)
and 973 Program (No.2007CB815402). R.M., C.L. and G.K. would like to 
thank the hospitality and stimulating atmosphere of the Aspen Center
for Physics where this work was initiated.  We also
thank Philip Best 
for providing the DR4 radio dataset before its publication, and for
useful comments on a draft of this manuscript.  Finally, we thank the
anonymous referee for the many helpful comments.

Funding for the SDSS and SDSS-II has been provided by the Alfred
P. Sloan Foundation, the Participating Institutions, the National
Science Foundation, the U.S. Department of Energy, the National
Aeronautics and Space Administration, the Japanese Monbukagakusho, the
Max Planck Society, and the Higher Education Funding Council for
England. The SDSS Web Site is http://www.sdss.org/. 

The SDSS is managed by the Astrophysical Research Consortium for the
Participating Institutions. The Participating Institutions are the
American Museum of Natural History, Astrophysical Institute Potsdam,
University of Basel, University of Cambridge, Case Western Reserve
University, University of Chicago, Drexel University, Fermilab, the
Institute for Advanced Study, the Japan Participation Group, Johns
Hopkins University, the Joint Institute for Nuclear Astrophysics, the
Kavli Institute for Particle Astrophysics and Cosmology, the Korean
Scientist Group, the Chinese Academy of Sciences (LAMOST), Los Alamos
National Laboratory, the Max-Planck-Institute for Astronomy (MPIA),
the Max-Planck-Institute for Astrophysics (MPA), New Mexico State
University, Ohio State University, University of Pittsburgh,
University of Portsmouth, Princeton University, the United States
Naval Observatory, and the University of Washington. 

\bibliography{cheng,rachel}

\bsp
\label{lastpage}

\end{document}